\documentclass[aip,pop,graphicx,preprint,floatfix]{revtex4-1}
\usepackage{graphicx}
\usepackage{dcolumn}
\usepackage{bm}

\usepackage[utf8]{inputenc}
\usepackage[T1]{fontenc}
\usepackage{mathptmx}
\usepackage{etoolbox}

\makeatletter
\def\@email#1#2{%
 \endgroup
 \patchcmd{\titleblock@produce}
  {\frontmatter@RRAPformat}
  {\frontmatter@RRAPformat{\produce@RRAP{*#1\href{mailto:#2}{#2}}}\frontmatter@RRAPformat}
  {}{}
}%
\makeatother
\begin{document}

\preprint{AIP/123-QED}

\title{Ion density waves driving the formation of filamentary dust structures}
\author{A. Mendoza}
\author{D. Jiménez Martí}
\author{L.S. Matthews}
\author{B. Rodríguez Saenz}
\affiliation{ 
Center for Astrophysics, Space Physics, and Engineering Research (CASPER) at Baylor University, Waco, TX, 76706, USA
}
\author{P. Hartmann}
\affiliation{ 
Center for Astrophysics, Space Physics, and Engineering Research (CASPER) at Baylor University, Waco, TX, 76706, USA
}
\affiliation{
Institute for Solid State Physics and Optics, Wigner Research Centre for Physics, P.O. Box 49,
H-1525 Budapest, Hungary
}
\author{E. Kostadinova}
\affiliation{Physics Department, Auburn University, Auburn, AL, 36849, USA}
\author{M. Rosenberg}
\affiliation{Department of Electrical and Computer Engineering, University of California at San Diego, La Jolla, CA, 92037, USA}
\author{T.W. Hyde}
\affiliation{ 
Center for Astrophysics, Space Physics, and Engineering Research (CASPER) at Baylor University, Waco, TX, 76706, USA
}
\email{Lorin\_Matthews@baylor.edu.}
\date{\today}

\begin{abstract}
The PlasmaKristall-4 experiment on the International Space Station allows for the study of the 3-dimensional interaction between plasma and dust particles. Previous simulations of the PK-4 environment have discovered fast moving ionization waves in the dc discharge \cite{hartmann2020}. These ionization waves vary the plasma parameters by up to an order of magnitude, which may affect the mechanisms responsible for the self-organization of chains seen in the PK-4 experiment. Here, we adapt a molecular dynamics simulation to employ temporally varying plasma conditions in order to investigate the effect on the dust charging and electrostatic potential. In order to describe the differences between the average of the plasma conditions and the time-varying plasma condition, we present a model to reproduce the potential that takes into account the negative  potential from the dust grain and the positive potential from the ion wake.
\end{abstract}

\maketitle

\section{\label{sec:intro}Introduction}
The underlying mechanisms leading to spontaneous structure formation comprise a fascinating area of study. Complex plasma systems (systems of ionized gas and micrometer-sized solid particles) are ideal for investigating this phenomenon since the dynamics of individual particles occur on easily accessible time and spatial scales. Experiments have provided insight into phenomena such as structural phase transitions \cite{dietz2018}, demixing \cite{dietz2017}, rogue wave generation \cite{tsai2016}, and non-equilibrium dynamics \cite{worner2012}. However, experiments on Earth are hampered by the fact that dust must be supported against the force of gravity, which causes many of the structures under investigation to be quasi-2D. 

The solid particles, often called dust particles, in the plasma tend to become negatively charged due to the greater frequency of collisions with electrons than ions in the plasma. In experiments conducted on Earth, the dust particles usually levitate in the electric field of the plasma sheath, compensating for the gravitational force \cite{chen2016}. The same electric field that levitates the dust particles causes ions to flow downward toward the lower electrode. This ion flow creates an ion wake downstream of the dust particles. As such, the ion wake is critical in understanding particle interactions and forming ordered dust structures. In order to investigate this wakefield, one dust grain must be downstream of another, which can be achieved by increasing the horizontal confinement force. On Earth, the wake force is directed primarily along the force of gravity, which makes it difficult to distinguish and investigate as it is much smaller than the gravitational force. One solution is to perform the experiment in microgravity, which effectively eliminates the gravitational force.

Since 2014, the PlasmaKristall-4 (PK-4) system on board the International Space Station has been used to conduct complex plasma experiments under microgravity \cite{dietz2018}, which enables the study of three-dimensional dust structures. A surprising effect in many of these experiments was the formation of aligned grains \cite{sutterlin2009}, which is thought to be driven by the electrostatic interaction between the charged dust grains and ion wakes. Previous experiments performed in the PK-4 have given insight into the dependence of dust structure formation on discharge current \cite{du2012} and phase transitions in dust crystal structure \cite{dietz2018}. However, the exact mechanisms responsible for the dust structures are still unknown and an active research area.

In the PK-4, the primary process thought to be responsible for the dust chains seen is the electrorheological (ER) effect. ER fluids commonly consist of colloidal particles immersed in a fluid of a different dielectric constant \cite{chen1992}. In an ER fluid, an externally applied electric field determines the interparticle interaction by polarizing the particles and introducing a dipole-dipole coupling, leading to string formation. A similar effect has been experimentally observed in complex plasmas, where an external, alternating electric field distorts the positive ion cloud around a negatively charged dust particle. The symmetric ion wakes are proposed to give rise to an attractive dipole-dipole interaction. \cite{ivlev2008}. However, recent simulations have suggested a reduced repulsion between dust grains is responsible for the chain strucutes \cite{joshi2023}.

A numerical simulation of the plasma conditions inside the PK-4 discharge using a two-dimensional particle-in-cell (PIC) model with Monte Carlo collisions (MCC) \cite{hartmann2020} revealed the presence of ionization waves in the dc discharge. These ionization waves are regions of high particle density moving through the column with phase velocities of 500-1200 m/s. The peak in the axial electric field is approximately 20 times larger than the axial electric field between the ionization waves. In the region of the homogeneous plasma column, the plasma conditions change on a microsecond timescale while the dust responds on a timescale of milliseconds. Thus, it might be assumed that the time-averaged plasma conditions could be used to calculate the dust charging and particle interaction \cite{matthews2021}, \cite{vermillion2022}. 

However, previous simulations of the dust dynamics and ion wakes found that time-averaged plasma temperatures and axial electric field lead to weakly ordered string structures \cite{matthews2021}. An increased axial electric field was needed to achieve a chain of particles with order similar to the string-like structures observed in the PK-4 experiment \cite{matthews2021}. A follow-up study \cite{vermillion2022} examined both the dust charging behavior and wake formation using the plasma conditions present at different points in an ionization wave. It was shown that the dust charge in the varying plasma environment should be less than the charge obtained using the averaged plasma conditions. These smaller charge is a result of the delayed charging and discharging of the dust grain as the ionization wave passes. Together, these simulations showed that varying conditions within the ionization waves play an essential role in the structural order of the dust cloud observed in the PK-4 experiment.

This work investigates how temporal variations in the plasma conditions affect dust charging and ion wake formation. An overview of the PK-4 experiment and a numerical model of the plasma conditions within the dc discharge are given in Section 2. Section 3 describes the modeling of the interaction of ions and dust within the discharge, using both constant and time-varying plasma conditions. The results of these simulations are presented in Section 3, with an analysis of the differences in the wake characteristics in Section 4.

\section{\label{sec:background}Background}

\subsection{PK-4}

\begin{figure}[ht!]
    \centering
    \includegraphics[width=3.37in]{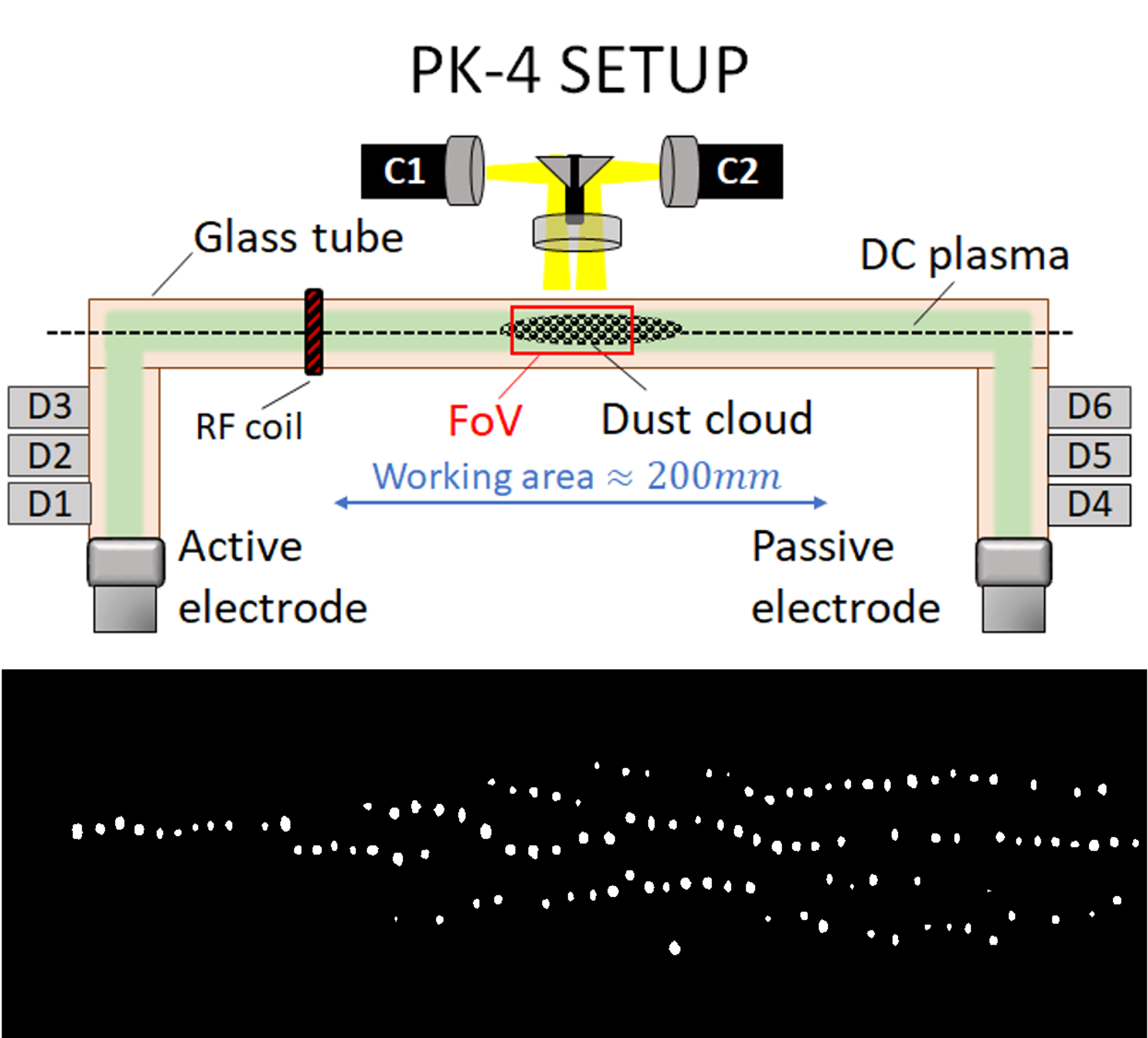}
    \caption{Schematic of the PK-4 experiment showing the $\pi$-shaped configuration of the discharge tubes, the relative location of the particle observation cameras (C1 and C2), dust shakers (D1-D6), and the field of view in the working area (FoV). The bottom half of the figure shows dust strings in the PK-4.}
    \label{fig:pk4}
\end{figure}

The Plasmakristall-4 (PK-4) is a microgravity complex plasma laboratory installed in the Columbus module of the ISS. The PK-4 is capable of generating a dc or radio frequency (rf) plasma discharge using argon or neon gases, with the option of imposing polarity switching in the axial dc electric field. The plasma chamber is comprised of three 30-mm-diameter glass tubes connected in a $\pi$-shaped configuration with a 200-mm working area located in the central part of the main tube. Two particle observation cameras visualize dust clouds, and a plasma glow observation camera is used to monitor the discharge. An illumination laser and manipulation laser are positioned at either end of the main tube [not shown in Fig. \ref{fig:pk4}], and three dust shakers for dispensing micrometer-sized grains are located on each side tube. A schematic of the PK- 4 experiment is shown in Fig. \ref{fig:pk4}.

Dust grains immersed in the PK-4 plasma are subject to acceleration by the axial electric field. In order to trap dust grains in the central field of view for video observation, the polarity of the dc electric field is alternated uniformly at a sufficiently high frequency that the dust grains are unable to respond to the changing field direction, typically 500 Hz (while dust response is commonly in the range 10 Hz – 100 Hz). However, since the less massive ions can respond to changing conditions at timescales on the order of microseconds, the changing electric field direction causes the ions to flow past the relatively stationary dust grains during both phases of the polarity switching. Positively charged ions flowing past a negatively charged dust grain will be deflected, forming a region of enhanced ion density downstream from the dust grain, known as an ion wake. In addition to trapping the dust grains within the field of view, it has also been observed that applying uniform polarity switching of the axial electric field can also cause the dust grains to align in long filamentary chain structures \cite{ivlev2011}, like those shown in Fig \ref{fig:pk4}.
 
\subsection{Gas Discharge Modeling}

\begin{figure}[ht!]
    \centering
    \includegraphics[width=3.37in]{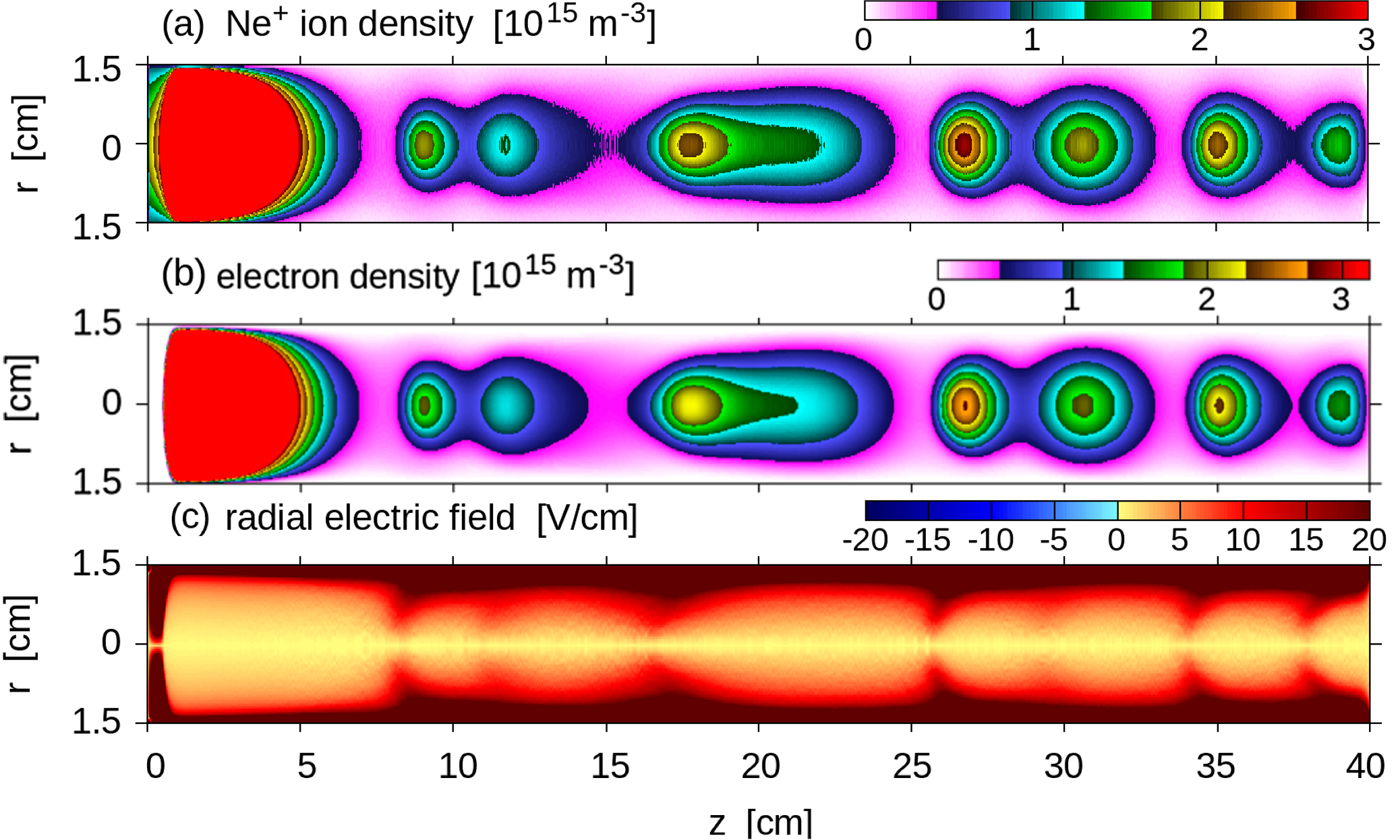}
    \caption{Computed spatial distributions of plasma parameters: (a) Ne ion density and (b) axial electric field (where positive indicates in the direction of increasing z) at p = 60 Pa and I = 2 mA with the cathode at z = 0. Data gathered over 1 $\mu$s.}
    \label{fig:picmcc}
\end{figure}

The plasma in the PK-4 dc discharge (in the absence of dust) was modeled using a 2D Particle-in-Cell with Monte Carlo collisions (PIC-MCC) simulation. The PIC code assumed cylindrical symmetry and was used to mimic the conditions inside the PK4 experiment. The operating conditions were neon gas at $T_g = 300$ K, with a pressure $p = 60$ Pa, current of $I = 2 mA$ and a voltage of $\phi = -770$ V on the powered electrodes. The Monte Carlo method allows for the inclusion of collisions between the charged particles and neutral background gas. The charged particle species can interact and charge the electrode surfaces and the glass cylinder. The electric field is self-consistently derived through Poisson's equation using the boundary conditions at both the electrodes and walls of the glass cylinder.

The PIC-MCC simulation showed quasiperiodic variations in the plasma (Fig. \ref{fig:picmcc}a). These variations are known as ionization waves and are caused by variations in the ionization rate, driven by the interaction of electron dynamics (Fig. \ref{fig:picmcc}b) and electric fields (Fig. \ref{fig:picmcc}c) in the plasma, along with contributions from plasma instabilities and wave-particle interactions. The presence of ionization waves was confirmed by ground-based experiments using the PK-4 BU experiment, where a high-speed CCD camera was used to measure the light emission sequence \cite{schmidt2020}.

In addition to the naturally occurring ionization waves seen in the positive plasma column, the dc switching of the electrodes causes a substantial variation in the plasma conditions. However, this paper is only concerned with the effects of ionization waves on dust. A full description of the conditions under these parameters can be found in \cite{hartmann2020}.

\section{\label{sec:numerical}Numerical Simulation of Ions and Dust}

We are interested in determining the effects of ionization waves on dust charging and ion wake formation. Accordingly, we model the dynamics of ions flowing past charged dust grains, driven by the dc column's axial electric field and ionization waves. The multiscale model Dynamic Response of Ions and Dust (DRIAD) has previously been used to model ion dynamics and the charging and dynamics of the dust in a flowing plasma on their individual timescales \cite{matthews2021}, \cite{vermillion2022}, \cite{matthews2020}. This study adapted the model in order to include time-varying boundary conditions to account for the ionization waves observed in the PK-4 PIC model.

\subsection{DRIAD}
The dust grains are held fixed within the cylindrical simulation region to study the dust charging and ion wake formation. Therefore, the equation of motion of the dust grains is not used and not discussed here. Ions flow through the cylinder driven by the axial electric field within the dc discharge. The ion-ion forces, ion-dust forces, axial electric field, the confining force at the boundaries, and ion-neutral collisions when taken together determine the motion of an ion, and can be written as:
\begin{equation}
    m_i \vec{\ddot{r}} = \vec{F}_{ij} + \vec{F}_{id} + \vec{F}_{E}(z) + \vec{F}(r,z)+ \vec{F}_{in}
\end{equation}
where $F_{ij}$ is the Yukawa force between ion $i$ and ion $j$:
\begin{equation}
    \vec{F_{ij}} = \sum_{i\neq j} \frac{q_iq_j}{4\pi\epsilon_o(r_{ij}+r_o)^3}(1+\frac{r_{ij}}{\lambda_{De}})exp\left(\frac{-r_{ij}}{\lambda_{De}}\right)\vec{r_{ij}}
\end{equation}
where $r_{ij}$ is the distance between ions $i$ and $j$, $q_i$ and $q_j$ are the charge of ion $i$ and $j$, $r_o$ is a softening radius to avoid singularities, and $\lambda_{De} = \sqrt{\frac{\epsilon_ok_bT_e}{n_e q_e^2}}$ is the electron Debye length where $T_e$ is the electron temperature, and $n_e$ is the electron density.

In the above, $F_{id}$ is the force due to the Coulomb interaction between the ions and dust particle, with charge $Q_d$

\begin{equation}
 \vec{F_{id}} = \sum_{d} {\frac{q_iQ_d}{4\pi\epsilon_o(r_{id}+r_o)^2}\vec{r_{id}}}   
\end{equation}
where $r_{id}$ is the distance between an ion $i$ and dust particle $d$.

$\vec{F_e}(z) = q_i\vec{E}(z)$ is the force due to the axial electric field, which causes the ions to flow through the cylindrical simulation region. $F_{in}$ is the force due to ion-neutral collisions found using the null-collision method, and $\vec{F}(r,z)$ represents the force from the ions outside of the simulation region, where $\vec{E_b}(r,z)$ is the electric field within a cylindrical cavity found by assuming a uniform distribution of ions within a cylinder and subtracting a constant potential. Ions that leave the simulation region or are absorbed by the dust grains are reinserted on the boundaries, consistent with the direction of ion flow.

\subsection{Time-varying boundary conditions}

\begin{figure}[ht!]
    \centering
    \includegraphics[width=3.37in]{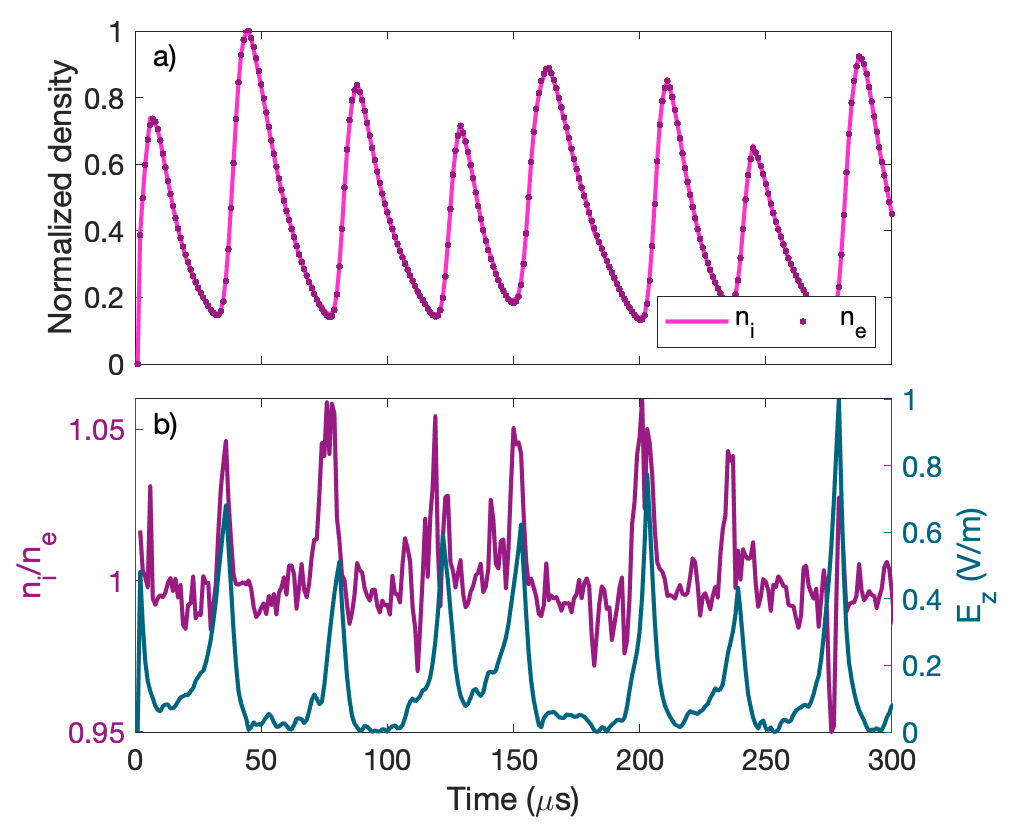}
    \caption{a) Variations in electron and ion density in time. Values have been normalized by their respective maximums. b) The electron and ion density ratio at a point in the center of the discharge column (left axis, purple) and the resultant axial electric field (right axis, teal). Data obtained from the PIC-MCC model for 60 Pa. All data has been normalized by their maximum value.}
    \label{fig:den_e}
\end{figure}

The DRIAD code was revised to include time-varying boundary conditions based on the results of the PIC-MCC simulation. To include temporally evolving plasma conditions, a lookup table is generated at the beginning of the simulation, with boundary conditions for each plasma condition at one microsecond intervals.

The PIC-MCC results show that the electron and ion densities vary periodically, with peaks appearing approximately every 50 $\mu s$ (Fig. \ref{fig:den_e}a). The electron and ion number densities are approximately equal; however, as each ionization wave passes, there is approximately a 5$\%$ difference in the electron and ion densities, as seen in the time evolution of the ratio $n_i/n_e$ (Fig. \ref{fig:den_e}b, purple line). The local difference in electron and ion densities results in strong axial electric fields, which slightly lag the peaks in $n_i/n_e$ (Fig. \ref{fig:den_e}a). 
  
\begin{figure}[ht!]
    \centering
    \includegraphics[width=3.37in]{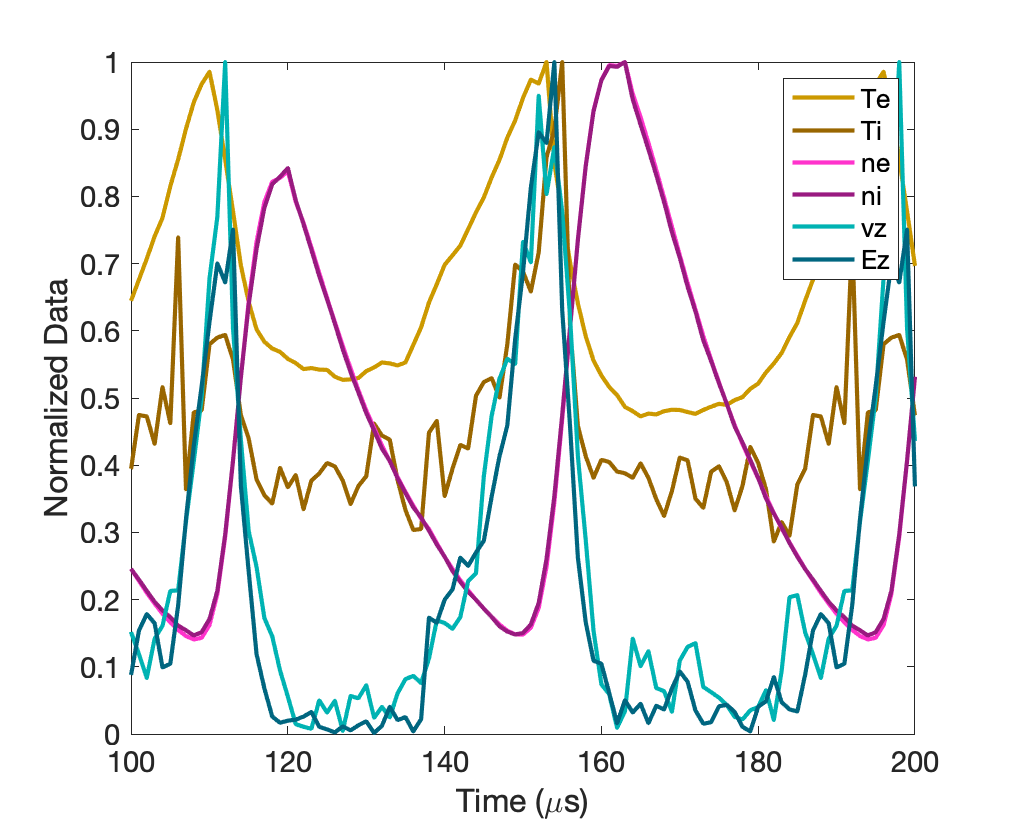}
    \caption{Representative data for time-varying plasma parameters at 60 Pa. Each value has been normalized by their respective maxima.}
    \label{fig:params}
\end{figure}

To include these effects in the DRIAD simulation, an 86 $\mu s$ time segment was selected from the complete PIC simulation data, representative of the recurring ionization waves. The data, output every 1 $\mu s$, was smoothed using a 3 $\mu s$ running average over each plasma condition. Fig. \ref{fig:params} shows an example of the repeating pattern for the electron and ion temperatures (gold), electron and ion densities (pink), and ion flow velocity and axial electric field (blue) for 60 Pa neon gas. The simulated cylinder had a height of 5420 $\mu$m and a 903 $\mu$m radius. The large simulation region ensured that the boundaries were sufficiently far from the dust grains to minimize the influence of computational effects from the boundaries. The dust grains were 1.69 $\mu$m in radius and placed 250 $\mu$m apart to match experimental conditions at 60 Pa.

Average plasma parameters are given in Table \ref{table:1}, where the average is taken over the entire 450 $\mu$s of simulated plasma conditions from the PIC-MCC simulation (CONST) and over the repeating 86 $\mu$s segment (EVOL). Due to the variation in the plasma parameters in the EVOL case, the electron Debye length varies from 300 $\mu$m to 1000 $\mu$m, but the average Debye length of the EVOL case is comparable to the CONST case, as shown in Table\ref{table:1} .

\begin{figure}[ht]
    \centering
    \includegraphics[width=3.37in]{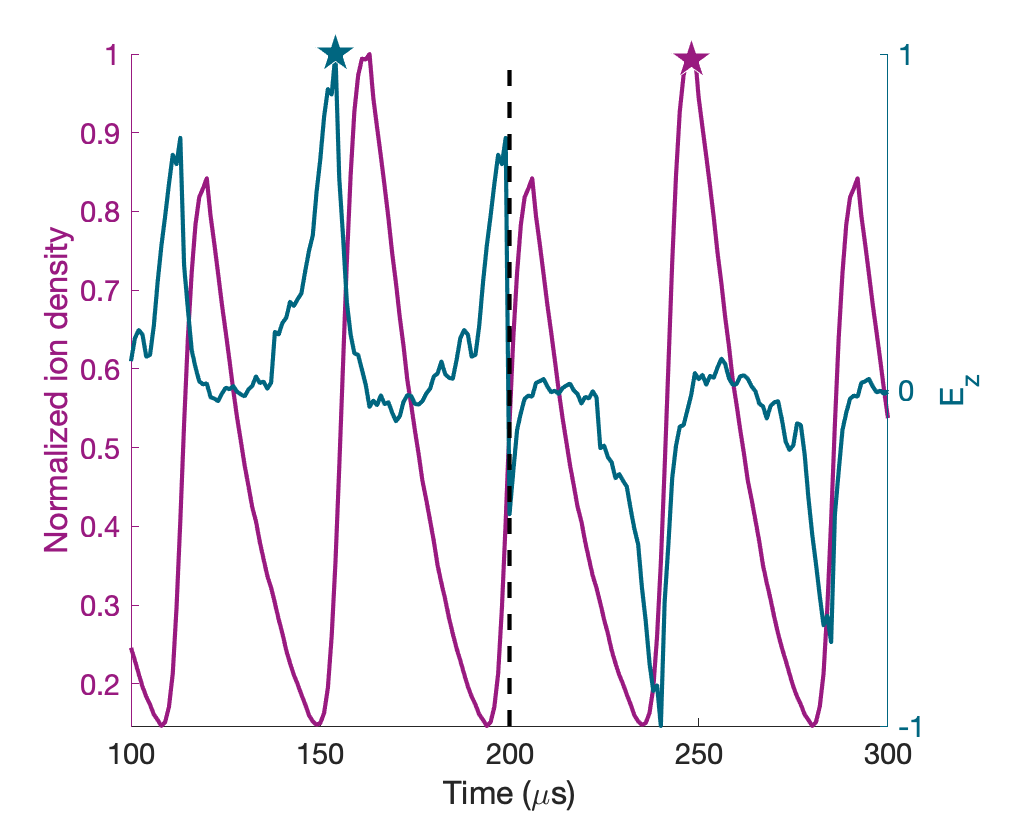}
    \caption{The time-varying ion density (purple, left axis) and axial electric field (teal, right axis). A dc electric field polarity switch is imposed on the axial electric field with a frequency of 5000 Hz. Normalized by their respective maximum values so that the peak density and electric field are 1. The teal star corresponds to a maximum in electric field with near minimum ion density, and the purple star corresponds to a maximum in ion density and the electric field is near minimum.}
    \label{fig:time-varying field and density}
\end{figure}

For computational expediency, we use super-ions, representing a cloud of ions with the same charge-to-mass ratio as a single ion, to model the ions. The number of super-ions in the simulation is held constant. However, the charge and mass of the newly injected ions are recalculated to match the density in the ionization waves. Fig. \ref{fig:params} shows the plasma conditions used to set boundary conditions in the DRIAD code, including the external electric field which provides confinement for the ions, the drift speed of injected ions, the charge on the injected ions, and the electron Debye length. The ion time step is $\Delta t_i = 1.0$ x $10^{-8}$ s; the ions are advanced for 100 timesteps before the plasma conditions change. Ions that leave the simulation boundary or are collected by dust grains are inserted at the boundary with a velocity determined by the current plasma condition. The density of ions in the simulation varies with the ionization waves.  The direction of the applied axial electric field (and subsequent drift velocity of the ions) is switched periodically to stabilize the dust cloud, as shown in Fig. \ref{fig:time-varying field and density}. We use a dc switch with a 5000 Hz frequency for computational efficiency; in the PK-4 experiment, the dc switching frequency is usually set to 500 Hz.

\begin{table}
    \caption{\label{table:1}Average plasma parameters for constant (CONST) and evolving (EVOL) plasma conditions.}
    \begin{ruledtabular}
    \begin{tabular}{lcc}
    Average Plasma Parameters & CONST & EVOL\\
    \hline
    $T_e$ (eV) & 4.3 & 4.3 \\
    $T_i$ (eV) & 0.041 & 0.043 \\
    $n_{io}$ ($10^{15}/m^3$) & 1.2 & 1.3\\
    $v_z$ (m/s) & $-$110 & $-$110\\
    $E_z$ (V/m) & 200 & 200\\
    $\lambda_{De}$ ($\mu$m) & 450  & 440\\
    $\lambda_{Di}$ ($\mu$m) & 44  & 43\\
    \end{tabular}
    \end{ruledtabular}
\end{table}

\section{Results}
The total simulation was run for 700 $\mu s$, covering three complete polarity switching cycles after the system comes to equilibrium. The results for dust charging, ion density, and electrostatic potential near the dust grain are compared for the two cases: CONST, using the average plasma conditions, and EVOL, where the plasma conditions vary and the average is taken. In the EVOL case, we show how the charge varies in time and the evolution of the ion wake as an ionization wave passes. When the ion density is at a maximum, the electric field is near its minimum (purple star in Fig. 5). When the electric field is at a maximum, the ion density is near a minimum (teal star in Fig. 5). We also compare the results for the CONST and EVOL cases averaged over several complete polarity switching cycles.

\subsection{Isolated dust grain}

\begin{figure}[ht!]
    \centering
    \includegraphics[width=3.37in]{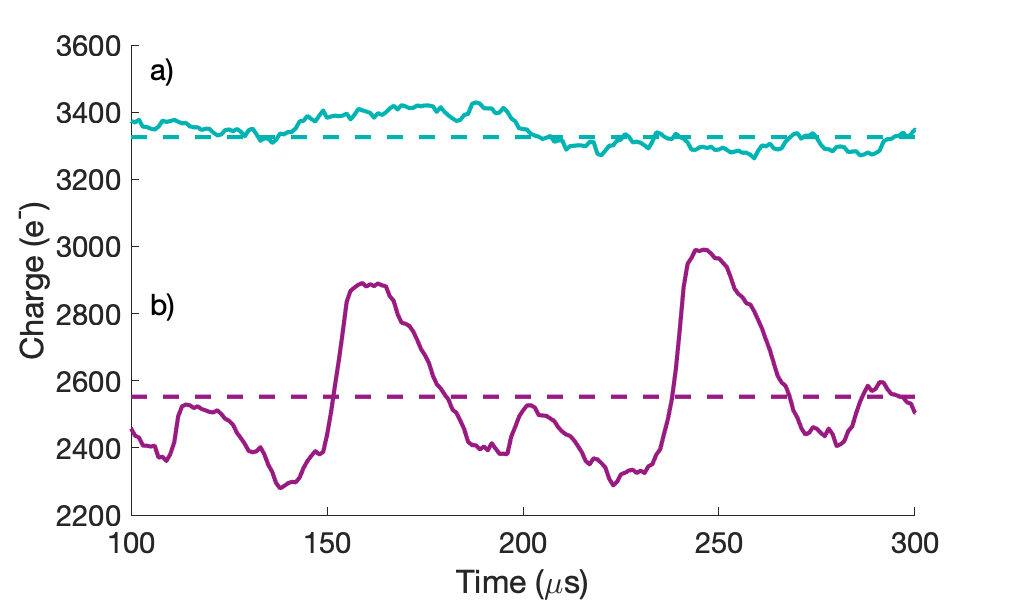}
    \caption{Dust charge for both (a) CONST and (b) EVOL conditions. The dashed line indicates the average charge.}
    \label{fig:onedust_charge}
\end{figure}

\begin{figure*}[ht!]
    \centering
    \includegraphics[width=6in]{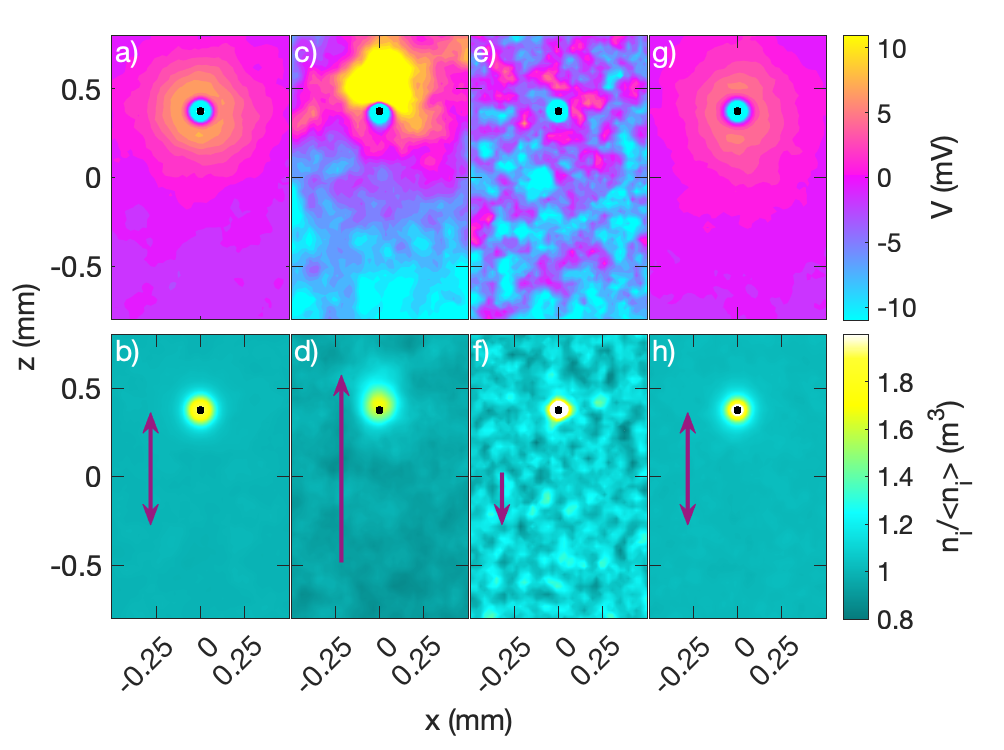}
    \caption{Electric potential (top) and ion density (bottom) for 60 Pa in the vicinity of a single dust grain. (a,b) corresponds to the CONST case, (c,d) correspond to the maximum electric field and near minimum ion density, corresponding to the teal star in Fig. \ref{fig:time-varying field and density}. (e, f) corresponds to the maximum ion density, corresponding to the purple star in Fig. \ref{fig:time-varying field and density}, and (g, h) corresponds to the time-averaged conditions. The ion density maps are normalized by the average ion density in the simulation for the given period. The ion flow for all cases is denoted by the purple arrows on the ion density plots, with longer arrows signifying stronger ion flow speeds.}
    \label{fig:potentialdensity_one}
    \end{figure*}
       
The variation in ion flow and ion density results in a smaller dust charge in the EVOL case compared to the CONST case. Figure \ref{fig:onedust_charge} depicts the variation in charging of the dust grain over time (solid line) as well as the equilibrium charge (dashed line) for both the CONST (a) and EVOL (b) cases. The equilibrium charge in the presence of ionization waves is approximately $76\%$ of the equilibrium charge of the average plasma conditions. While the average electron temperature in the CONST and EVOL cases are the same to two significant figures, it was previously shown that the charging time of a dust grain in the ionization wave is approximately 60 $\mu s$ \cite{vermillion2022}, while the time for the ionization wave to pass is approximately 20 $\mu$s. As the ionization wave is passing too fast, the dust charge in the EVOL case is lower than the CONST case. Although the ionization waves are happening on a timescale too fast for the dust to respond, the lower average dust charge leads to smaller interparticle forces.

Figure \ref{fig:potentialdensity_one} shows the electric potential and ion density for the case of a single dust grain. The CONST case (Fig. \ref{fig:potentialdensity_one} a) shows a strongly positive potential around the dust grain, though slightly asymmetric. When the electric field is at a maximum (Fig. \ref{fig:potentialdensity_one}c,d), the streaming ions are intensely focused downstream of the dust grains and produce a pronounced positive ion wake potential. When the ion density is high, and the electric field is at a near minimum, the low ion drift velocity results in a spherical cloud of ions (Fig. \ref{fig:potentialdensity_one}e,f).

The EVOL conditions averaged over several polarity cycles (Fig. \ref{fig:potentialdensity_one}g) shows a slight, asymmetric positive potential around the dust grain as expected. The ion density of the EVOL case (Fig. \ref{fig:potentialdensity_one}h) is much more concentrated around the dust grain than in the CONST case (Fig.\ref{fig:potentialdensity_one}b). However, the maximum ion density is greater in the EVOL case.

\subsection{Interacting dust grains}
To investigate the interaction between dust grains seen in chains of particles, the ion flow around a system of four particles separated by a distance of 250 $\mu$m was simulated.

\begin{figure}[ht!]
    \centering
 \includegraphics[width=3.37in]{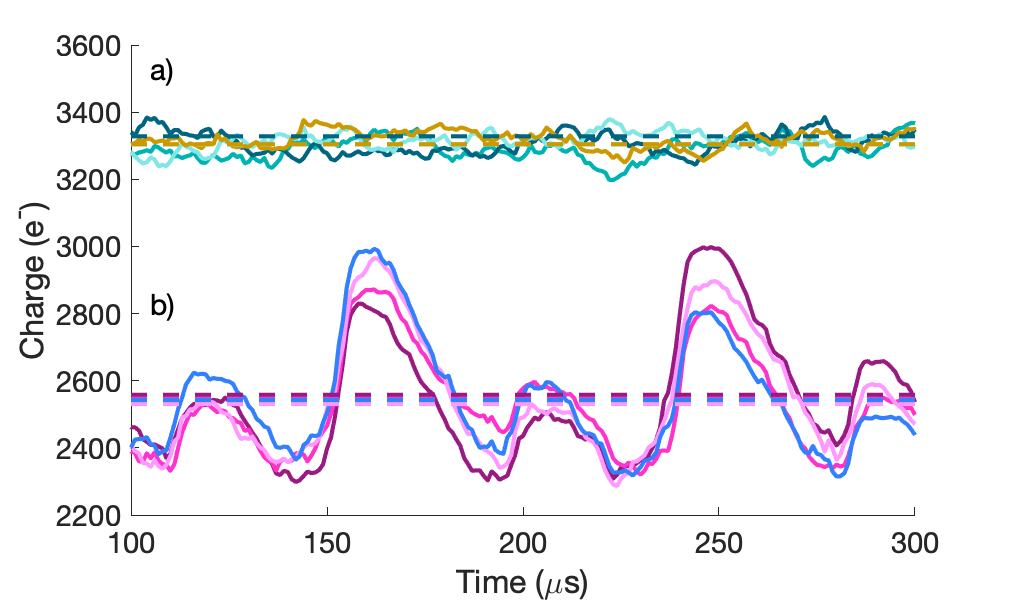}
    \caption{Variation in grain charge for (a) CONST and (b) EVOL cases.  The four dust grains have an interparticle separation of 250 $\mu$m. The horizontal dashed lines indicate the average grain charge. }
    \label{fig:charge4}
\end{figure}

Figure \ref{fig:charge4} shows the variation in the charge on the four grains for both the CONST (a) and EVOL cases (b). Similar to the case with a single grain, the grains have a higher average charge in the CONST case, with the charges on the grains in the EVOL case closely follow the periodicity in the ionization wave. The effects of a polarity switch are evident in Fig. \ref{fig:charge4}b, which occurs at 200 $\mu$s.

\begin{figure*}[ht!]
     \centering
    \includegraphics[width=6in]{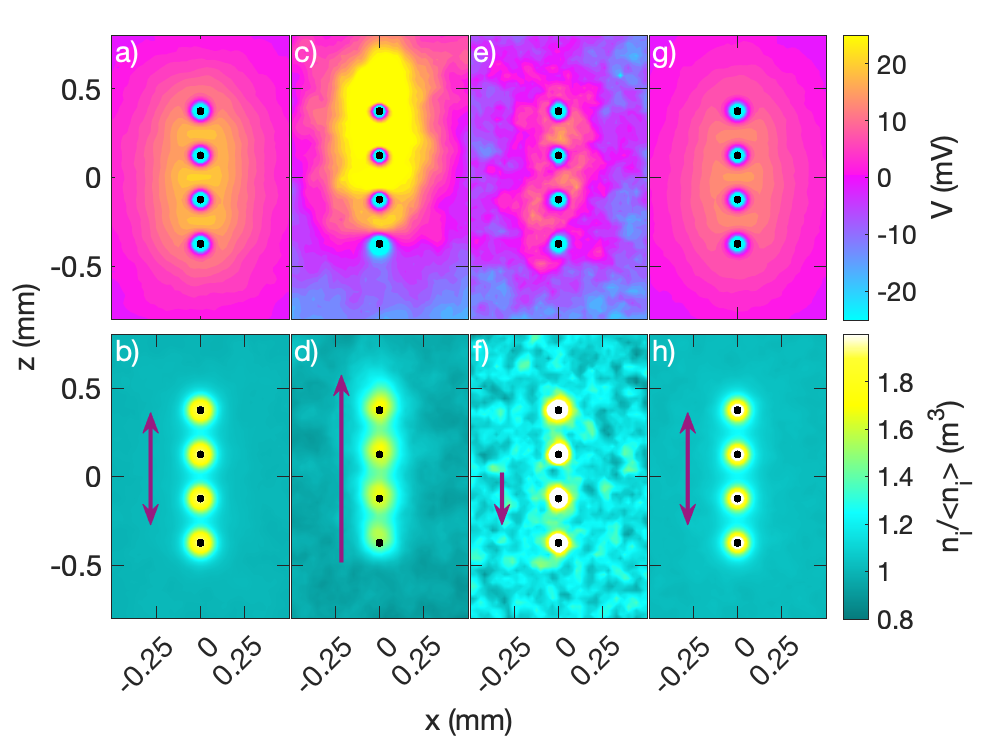}
    \caption{Electric potential (top) and ion density (bottom). 
    (a,b) corresponds to the CONST case, (c,d) correspond to the maximum electric field and near minimum ion density, corresponding to the teal star in Fig. \ref{fig:time-varying field and density}. In this case, the ion flow is directed in the positive z-direction. (e, f) corresponds to the maximum ion density, corresponding to the purple star in Fig. \ref{fig:time-varying field and density}, and (g, h) corresponds to the time-averaged conditions. The ion density maps are normalized by the average ion density in the simulation for the given period. The ion flow for all cases is denoted by the purple arrows on the ion density plots, with longer arrows signifying stronger ion flow speeds.}
    \label{fig:potentialdensity_4}
\end{figure*}

Figure \ref{fig:potentialdensity_4} depicts the electrostatic potential and the ion density for the four particles at two different instances (maximum axial electric field, maximum ion density) in the ionization wave, as well as the average of the time-varying and time-average plasma parameters.

Figure \ref{fig:potentialdensity_4}a, b displays the CONST conditions averaged over multiple polarity cycles. The positive ion wake potential (a) is very pronounced in the region of the dust grains. In addition, the region between the dust grains is strongly positive. The ion density (b) shows regions of ion concentration around each grain.

Figure \ref{fig:potentialdensity_4}c-h show results for the EVOL case. At the peak electric field (Fig. \ref{fig:potentialdensity_4}c,d) the ion clouds are displaced downstream of the grains, and there is an extended positive ion wake potential. When the plasma density is at a maximum with low ion drift velocity (Fig. \ref{fig:potentialdensity_4}e,f), the ion clouds about each grain are very dense but distinct from each other, and there is only a small positive potential between the dust grains. Results for the EVOL conditions averaged over several polarity switching cycles are shown Fig \ref{fig:potentialdensity_4} g,h.  The asymmetry in the ion clouds is clearly seen.  The positive ion wakes between the grains are less pronounced than in the CONST case, but that is to be expected since the smaller dust potential leads to less ion focusing.

\section{Discussion}
\subsection{Potential fitting}

\begin{figure}[ht!]
	\centering
	\includegraphics[width=3.37in]{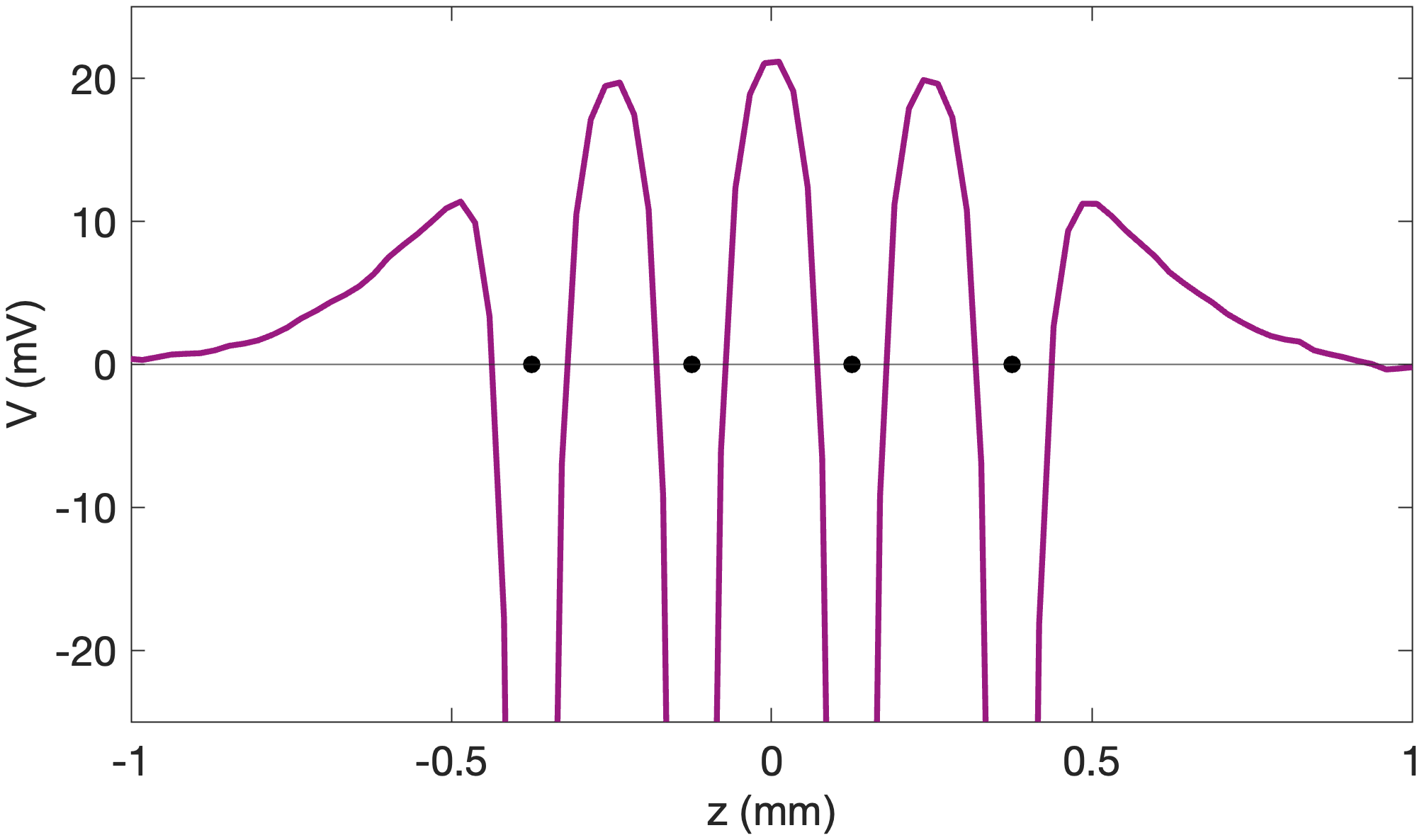}
	\caption{Potential at x=0, z for a system composed of four dust grains along the cylinder axis, separated by a distance $\Delta z=250$ $\mu m$. The dust grain locations are indicated by black dots.}
	\label{potential_example}	
\end{figure}

As shown, there are various differences in the charging and structure on the ion wakes for CONST and EVOL plasma conditions. Several key features of the electric potential can be extracted from the potential along the cylinder axis (x = 0, z) (Fig.\ref{potential_example}). The potential takes highly negative values near the dust grains and is positive in the region between dust grains. At long distances, the potential decays towards zero. This spatial behavior leads us to propose a model (\ref{potential_model}) to reproduce the potential distribution near a chain of dust particles and to quantify these differences. The proposed potential model is composed of a contribution from the negative dust grain, given by a screened Coulomb potential, and a positive contribution due to the ion wake, represented by an anisotropic Gaussian function centered at each dust grain. The data obtained from DRIAD shown in Fig. \ref{fig:potentialdensity_one}a, g for a single grain and \ref{fig:potentialdensity_4}a, g for a chain of four grains for CONST and EVOL plasma conditions were fit using the proposed potential model as shown in Eq. \ref{potential_model}.

\begin{equation}  
		\label{potential_model}
		V(x, z) = \sum_{i=1}^{N}\frac{Q_i}{4\pi\epsilon_0}\frac{e^{-\frac{|\vec{r}-\vec{r}_i|}{\lambda_i}}}{|\vec{r}-\vec{r}_i|} + A_{i}e^{-[\frac{(x-x_i)}{B_i}]^2}e^{-[\frac{(z-z_i)}{C_i}]^2}
\end{equation} 
where $Q_i$ and $\vec{r}_i = (x_i, z_i)$ are the charge and position vector of the $i$-th dust grain and the coefficients $A_i$, $B_i$, $C_i$ and $\lambda_i$ are all positive. The coefficient $A_i$ represent the maximum of the positive ion wake potential. Hence, a greater value of this coefficient means a more pronounced ion wake. The coefficients $B_i$ and $C_i$ determine the decay of this positive contribution in the x and z directions, such that a smaller value represents a faster decay. The isotropic shielding of the negative dust grain is represented by $\lambda_i$; a small coefficient indicates greater shielding. 

\begin{figure}[ht!]
	\centering
	\includegraphics[width=3.37in]{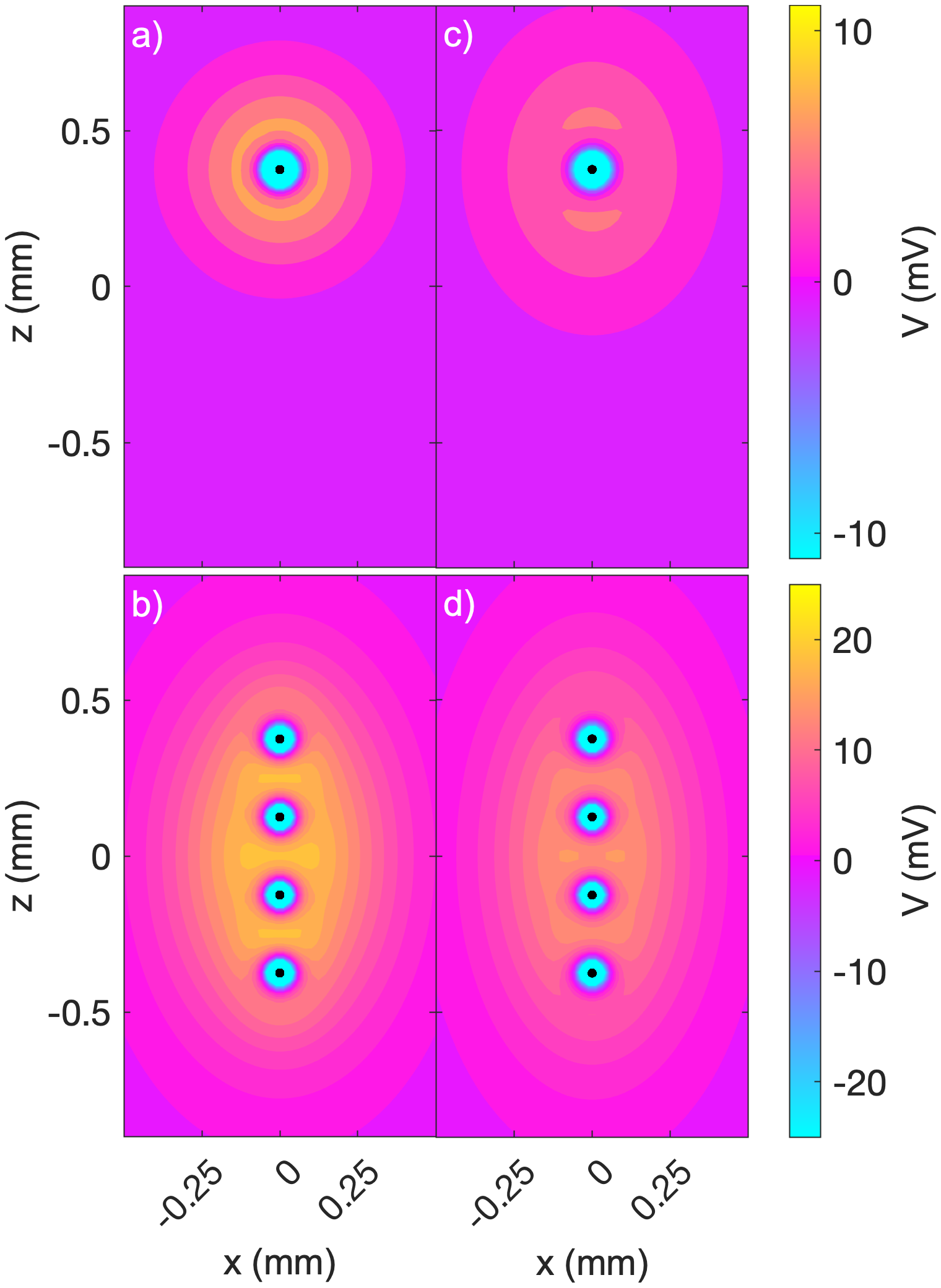}	\caption{Potential distribution obtained with the model in Eq. \ref{potential_model} for: a, b) CONST and c, d) EVOL conditions.}
	\label{potential_contour}
\end{figure}

\begin{figure}[ht!]
	\centering
	\includegraphics[width=3.37in]{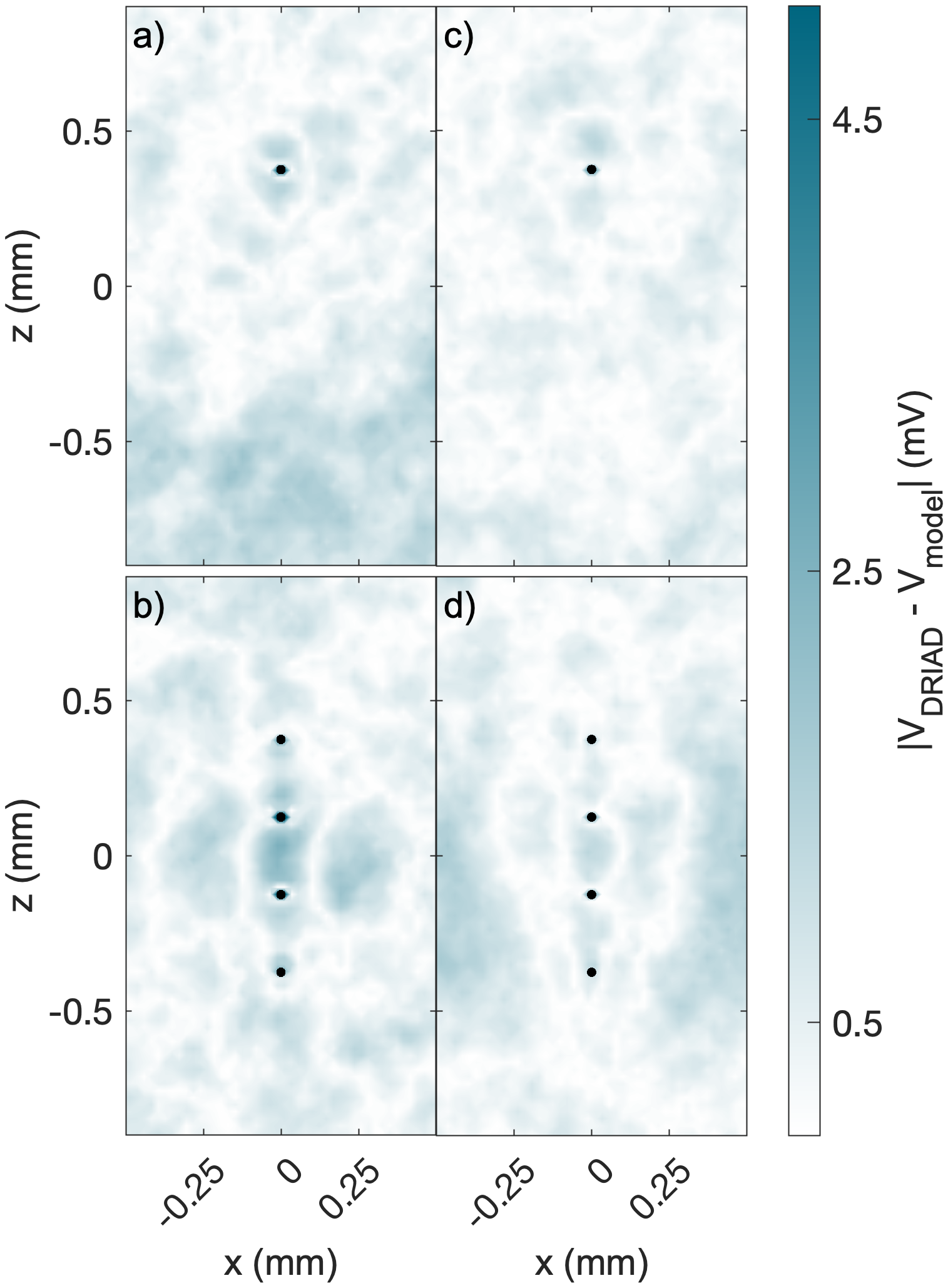}
	\caption{Difference between the potential distribution shown in Fig. \ref{potential_contour} obtained with the Gaussian ion wake model in Eq. \ref{potential_model} and the data from DRIAD for: a, b) CONST and c, d) EVOL conditions.}
	\label{potential_difference}
\end{figure}

The resulting potential distributions obtained using this model are shown in Fig. \ref{potential_contour}. The absolute value of the differences between the potential distribution obtained with the model and the data from DRIAD are shown in Fig. \ref{potential_difference}. Interestingly, the potential in the region near the dust grains is in better agreement for the EVOL plasma conditions than for the CONST plasma conditions since the magnitude of the difference is lower in this region.
	
\begin{figure}[ht!]
	\centering
	\includegraphics[width=3.37in]{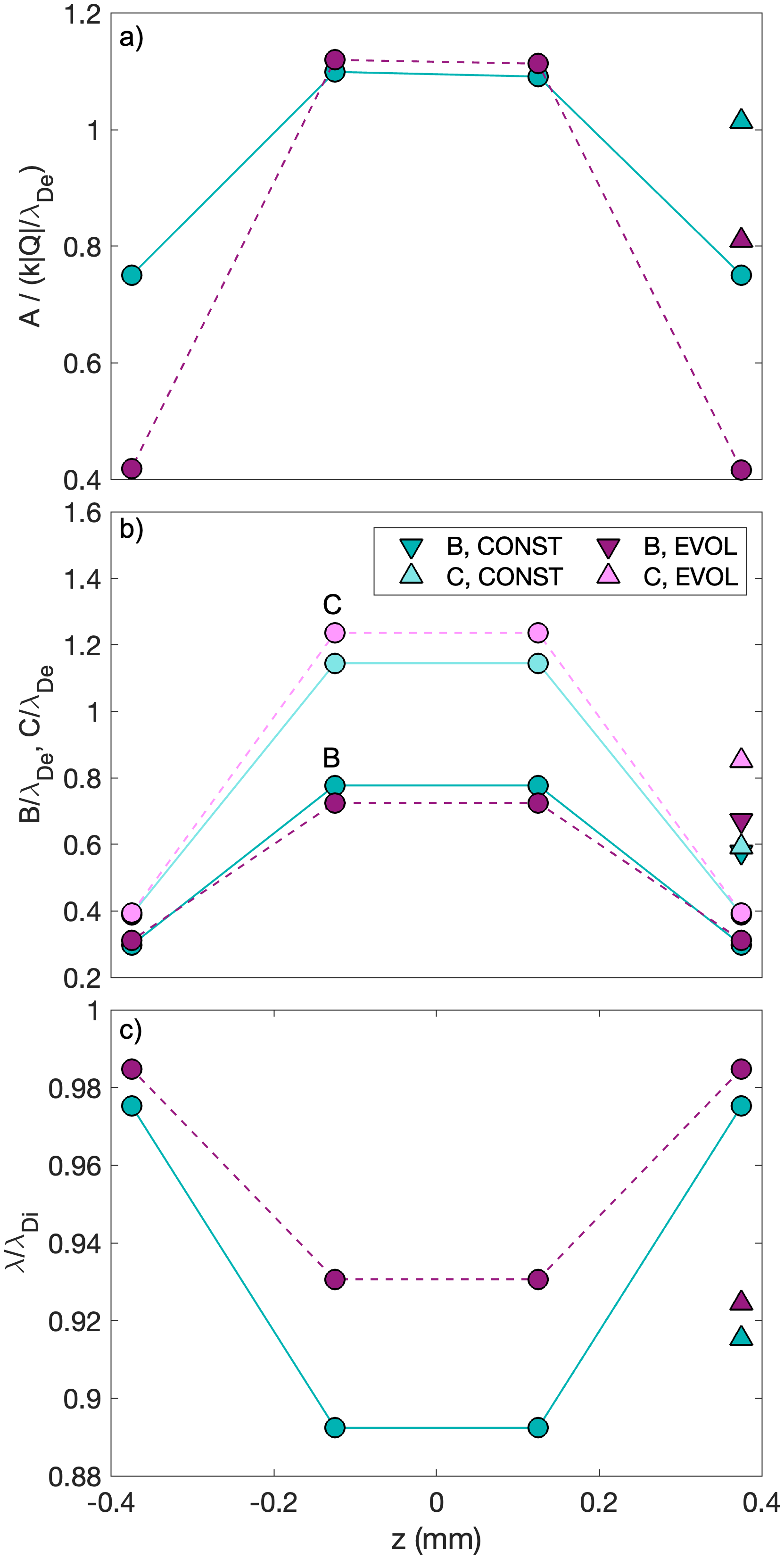}
	\caption{Normalized coefficients as a function of the dust position for CONST (teal circles with solid lines) and EVOL (pink circles with dashed lines) plasma conditions. Results for a single dust grain are indicated by triangles. All data are normalized by $\lambda_{De}$ or $\lambda_{Di}$ calculated for CONST conditions as listed in Table \ref{table:1}.}
	\label{coefficients}
\end{figure}

The coefficients obtained from the potential fitting and the average charge are shown in Tables \ref{coefficients_table_one} and \ref{coefficients_table_four}. The normalized coefficients comparing the results for the chain of four dust grains to those for an isolated dust grain are shown in Fig. \ref{coefficients}. As shown, there is a symmetry between the coefficients of the two inner and the two outer dust grains consistent with the symmetric positions of the dust grains in the simulation region.

\begin{table}[ht!]
    \caption{\label{coefficients_table_one}Average dust charge obtained from the DRIAD simulation and coefficients for the potential obtained from fitting the data with Eq. \ref{potential_model} for a single dust grain.}
    \begin{ruledtabular}
    \begin{tabular}{lcc}
                                &                  CONST                 &         EVOL                 \\	\hline
	$<Q>$ ($e^-$)	            &				   3330		             &		   2550					\\	
	 A (mV)	         	        &				   10.8			         &		   6.60				    \\	
	 B ($10^2\mu$m) 	        &				   2.60				     &		   3.03					\\	
	 C ($10^2\mu$m)	            &				   2.67				     &		   3.85					\\	
	$\lambda$ ($\mu$m)	        &				   39.9		             &		   40.3					\\	
    \end{tabular}
    \end{ruledtabular}
\end{table} 
 
\begin{table}[ht!]
	\caption{\label{coefficients_table_four}Average dust charge obtained from the DRIAD simulation and coefficients for the potential obtained from fitting the data with Eq. \ref{potential_model} for a chain of four dust grains.} 	
    \begin{ruledtabular}
	\begin{tabular}{lcccc}
				
												&      \multicolumn{2}{c}{Inner dust grains}            &      \multicolumn{2}{c}{Outer dust grains}       \\     \hline
										&	   CONST        &         EVOL        &	      CONST        &       EVOL         \\  
				$<Q>$ ($e^-$)	        &		3320		&		  2540		  &		   3300		   &	   2550			\\	 
				A (mV)	         	    &		11.6		&		  9.04		  &		   7.91	       &	   3.39 		\\	  
				B ($10^2\mu$m) 	        &		3.51		&		  3.27		  &		   1.34  	   &	   1.41  		\\	  
				C ($10^2\mu$m)	        &		5.17		&		  5.58		  &		   1.76	       &	   1.77		\\	  
				$\lambda$ ($\mu$m)	    &		38.9		&		  40.5	      &		   42.5	       &	   42.9		\\  			
				
	\end{tabular}
	\end{ruledtabular}
\end{table} 

As shown in Fig. \ref{coefficients}a, the maximum positive potential of the ion wake relative to the dust potential in the single dust case is smaller in the EVOL conditions than in the CONST conditions. Note that the values shown are normalized by the potential of the corresponding dust grain at a distance of one Debye length. The coefficients B and C of the single grain, Fig. \ref{coefficients}b, which represent the extent of the positive ion wake, are greater for the EVOL case than for the CONST case, which indicates a more relaxed wake for the EVOL case. The coefficients B (Fig. \ref{coefficients}b) of the outer dust grains are approximately equal for both types of conditions, whereas these coefficients are greater for the inner dust grains in the CONST conditions case, indicating a smaller wake in the horizontal direction for the EVOL case. This trend is reversed for the coefficients C (Fig. \ref{coefficients}b) where the largest coefficients are found for the inner dust grains in the EVOL plasma conditions case, which indicates a more extended wake in the axial direction. The coefficients B and C are on the order of the electron Debye length. The symmetric shielding of the dust, characterized by the length scale $\lambda$ (Fig. \ref{coefficients}c) is of the same order as the ion Debye length, consistent with shielding of the negative dust charge being provided by the ions. All values of $\lambda$ are greater for the EVOL plasma conditions than for CONST plasma conditions. This indicates less shielding of the negative dust grain potential.
 
The values of C/B for both conditions are shown in Table \ref{CtoBratio}. In the single dust grain case the ion wake is nearly symmetric for CONST plasma conditions ($C/B\approx1$), whereas the ion wake is elongated in the direction of the ion flow ($C/B>1$) for EVOL plasma conditions. The ion wakes are noticeably more elongated for EVOL plasma conditions than for CONST plasma conditions for the two inner dust grains on the four dust chain. However, this ratio is approximately the same for the outer dust grains under CONST and EVOL plasma conditions.
	
\begin{table}[ht!]
	\caption{\label{CtoBratio}Ratio $C/B$  for CONST and EVOL plasma conditions.} 		
	\begin{ruledtabular}
	\begin{tabular}{lcc}
			&	CONST   & EVOL \\ \hline
				Single dust grain 	           &				1.03				  &		    	1.27						\\
				Four dust chain (inner dust grains) 	 	  &				    1.47				 &		       1.71						\\
				Four dust chain (outer dust grains) 	      &				   1.31				     &		       1.26						\\
				
 \end{tabular}
 \end{ruledtabular}
\end{table}  

\section{Conclusion}
A numerical model has been presented that shows the effect of the microsecond plasma variations on the charging and dynamics of embedded micron-sized dust grains, which typically have a dynamic response time on the millisecond time scale. A PIC-MCC model of the PK-4 experiment shows that the plasma conditions rapidly change as ionization waves travel through the discharge tube. These evolving conditions were applied to a molecular dynamics simulation of ions flowing past dust grains to study the dust charging and ion wake formation.

The model reveals that the fast plasma variations significantly affect the time-averaged charge of the dust grains and the magnitude of the ion wakes. In the EVOL case, the dust charge is smaller, and the ion wakes are more elongated, which may help explain the linear chains observed to form under some operating conditions in the PK-4 experiment.

The effects of the ionization waves help the alignment of dust grains in two ways: first, the average particle charge is reduced, decreasing the repulsion between grains. In the presence of ionization waves, the dust charge varies from 76-96$\%$ of the charge when using constant conditions. At the same time, the strongly focused ions lead to a positive potential along the direction of ion flow, which reduces the repulsion and aids in particle alignment.

This work focused on plasma variations due to the ionization waves commonly found in dc discharge tubes. Previous work has shown that the plasma is extinguished and must be reignited over approximately 200 $\mu$s \cite{hartmann2020}. This is expected to have further significant effects on the dust, which may explain some of the anomalous behavior of the dust. The analytic form of the ion wake, which is an improvement over other forms of the potential that do not match the actual behavior of the ions \cite{ivlev2010}, is in agreement with other models \cite{joshi2023} and is being applied to particles with varying interparticle distances in order to be used in one-component models of dust grains immersed in plasma.
\nocite{*}

\begin{acknowledgments}
The authors gratefully acknowledge support from the U.S. Department of Energy, Office of Science, Office of Fusion Energy Sciences under Award No. DE-SC0021334, and the National Science Foundation under Grant Nos. 2308743, 2308742, and 2148653. The authors also gratefully acknowledge Bradley Andrew for his help with the model used in the potential fitting.
\end{acknowledgments}

\section*{Data Availability Statement}
The data that support the findings of this study are available from the corresponding author upon reasonable request.

\bibliography{ionizationwavesbib}

\begin{thebibliography}{19}%
\makeatletter
\providecommand \@ifxundefined [1]{%
 \@ifx{#1\undefined}
}%
\providecommand \@ifnum [1]{%
 \ifnum #1\expandafter \@firstoftwo
 \else \expandafter \@secondoftwo
 \fi
}%
\providecommand \@ifx [1]{%
 \ifx #1\expandafter \@firstoftwo
 \else \expandafter \@secondoftwo
 \fi
}%
\providecommand \natexlab [1]{#1}%
\providecommand \enquote  [1]{``#1''}%
\providecommand \bibnamefont  [1]{#1}%
\providecommand \bibfnamefont [1]{#1}%
\providecommand \citenamefont [1]{#1}%
\providecommand \href@noop [0]{\@secondoftwo}%
\providecommand \href [0]{\begingroup \@sanitize@url \@href}%
\providecommand \@href[1]{\@@startlink{#1}\@@href}%
\providecommand \@@href[1]{\endgroup#1\@@endlink}%
\providecommand \@sanitize@url [0]{\catcode `\\12\catcode `\$12\catcode `\&12\catcode `\#12\catcode `\^12\catcode `\_12\catcode `\%12\relax}%
\providecommand \@@startlink[1]{}%
\providecommand \@@endlink[0]{}%
\providecommand \url  [0]{\begingroup\@sanitize@url \@url }%
\providecommand \@url [1]{\endgroup\@href {#1}{\urlprefix }}%
\providecommand \urlprefix  [0]{URL }%
\providecommand \Eprint [0]{\href }%
\providecommand \doibase [0]{http://dx.doi.org/}%
\providecommand \selectlanguage [0]{\@gobble}%
\providecommand \bibinfo  [0]{\@secondoftwo}%
\providecommand \bibfield  [0]{\@secondoftwo}%
\providecommand \translation [1]{[#1]}%
\providecommand \BibitemOpen [0]{}%
\providecommand \bibitemStop [0]{}%
\providecommand \bibitemNoStop [0]{.\EOS\space}%
\providecommand \EOS [0]{\spacefactor3000\relax}%
\providecommand \BibitemShut  [1]{\csname bibitem#1\endcsname}%
\let\auto@bib@innerbib\@empty
\bibitem [{\citenamefont {Hartmann}\ \emph {et~al.}(2020)\citenamefont {Hartmann}, \citenamefont {Rosenberg}, \citenamefont {Juhasz}, \citenamefont {Matthews}, \citenamefont {Sanford}, \citenamefont {Vermillion}, \citenamefont {Carmona-Reyes},\ and\ \citenamefont {Hyde}}]{hartmann2020}%
  \BibitemOpen
  \bibfield  {author} {\bibinfo {author} {\bibfnamefont {P.}~\bibnamefont {Hartmann}}, \bibinfo {author} {\bibfnamefont {M.}~\bibnamefont {Rosenberg}}, \bibinfo {author} {\bibfnamefont {Z.}~\bibnamefont {Juhasz}}, \bibinfo {author} {\bibfnamefont {L.~S.}\ \bibnamefont {Matthews}}, \bibinfo {author} {\bibfnamefont {D.~L.}\ \bibnamefont {Sanford}}, \bibinfo {author} {\bibfnamefont {K.}~\bibnamefont {Vermillion}}, \bibinfo {author} {\bibfnamefont {J.}~\bibnamefont {Carmona-Reyes}}, \ and\ \bibinfo {author} {\bibfnamefont {T.~W.}\ \bibnamefont {Hyde}},\ }\href {\doibase 10.1088/1361-6595/abb955} {\bibfield  {journal} {\bibinfo  {journal} {Plasma Sources Science and Technology}\ }\textbf {\bibinfo {volume} {29}},\ \bibinfo {pages} {115014} (\bibinfo {year} {2020})}\BibitemShut {NoStop}%
\bibitem [{\citenamefont {Dietz}\ \emph {et~al.}(2018{\natexlab{a}})\citenamefont {Dietz}, \citenamefont {Bergert}, \citenamefont {Steinmüller}, \citenamefont {Kretschmer}, \citenamefont {Mitic},\ and\ \citenamefont {Thoma}}]{dietz2018}%
  \BibitemOpen
  \bibfield  {author} {\bibinfo {author} {\bibfnamefont {C.}~\bibnamefont {Dietz}}, \bibinfo {author} {\bibfnamefont {R.}~\bibnamefont {Bergert}}, \bibinfo {author} {\bibfnamefont {B.}~\bibnamefont {Steinmüller}}, \bibinfo {author} {\bibfnamefont {M.}~\bibnamefont {Kretschmer}}, \bibinfo {author} {\bibfnamefont {S.}~\bibnamefont {Mitic}}, \ and\ \bibinfo {author} {\bibfnamefont {M.~H.}\ \bibnamefont {Thoma}},\ }\href {\doibase 10.1103/PhysRevE.97.043203} {\bibfield  {journal} {\bibinfo  {journal} {Physical Review E}\ }\textbf {\bibinfo {volume} {97}},\ \bibinfo {pages} {043203} (\bibinfo {year} {2018}{\natexlab{a}})}\BibitemShut {NoStop}%
\bibitem [{\citenamefont {Dietz}\ \emph {et~al.}(2018{\natexlab{b}})\citenamefont {Dietz}, \citenamefont {Kretschmer}, \citenamefont {Steinmüller},\ and\ \citenamefont {Thoma}}]{dietz2017}%
  \BibitemOpen
  \bibfield  {author} {\bibinfo {author} {\bibfnamefont {C.}~\bibnamefont {Dietz}}, \bibinfo {author} {\bibfnamefont {M.}~\bibnamefont {Kretschmer}}, \bibinfo {author} {\bibfnamefont {B.}~\bibnamefont {Steinmüller}}, \ and\ \bibinfo {author} {\bibfnamefont {M.}~\bibnamefont {Thoma}},\ }\href {\doibase 10.1002/ctpp.201700055} {\bibfield  {journal} {\bibinfo  {journal} {Contributions to Plasma Physics}\ }\textbf {\bibinfo {volume} {58}},\ \bibinfo {pages} {21} (\bibinfo {year} {2018}{\natexlab{b}})}\BibitemShut {NoStop}%
\bibitem [{\citenamefont {Tsai}\ \emph {et~al.}(2016)\citenamefont {Tsai}, \citenamefont {Tsai},\ and\ \citenamefont {I}}]{tsai2016}%
  \BibitemOpen
  \bibfield  {author} {\bibinfo {author} {\bibfnamefont {Y.-Y.}\ \bibnamefont {Tsai}}, \bibinfo {author} {\bibfnamefont {J.-Y.}\ \bibnamefont {Tsai}}, \ and\ \bibinfo {author} {\bibfnamefont {L.}~\bibnamefont {I}},\ }\href {\doibase 10.1038/nphys3669} {\bibfield  {journal} {\bibinfo  {journal} {Nature Physics}\ }\textbf {\bibinfo {volume} {12}},\ \bibinfo {pages} {573} (\bibinfo {year} {2016})}\BibitemShut {NoStop}%
\bibitem [{\citenamefont {Wörner}\ \emph {et~al.}(2012)\citenamefont {Wörner}, \citenamefont {Räth}, \citenamefont {Nosenko}, \citenamefont {Zhdanov}, \citenamefont {Thomas}, \citenamefont {Morfill}, \citenamefont {Schablinski},\ and\ \citenamefont {Block}}]{worner2012}%
  \BibitemOpen
  \bibfield  {author} {\bibinfo {author} {\bibfnamefont {L.}~\bibnamefont {Wörner}}, \bibinfo {author} {\bibfnamefont {C.}~\bibnamefont {Räth}}, \bibinfo {author} {\bibfnamefont {V.}~\bibnamefont {Nosenko}}, \bibinfo {author} {\bibfnamefont {S.~K.}\ \bibnamefont {Zhdanov}}, \bibinfo {author} {\bibfnamefont {H.~M.}\ \bibnamefont {Thomas}}, \bibinfo {author} {\bibfnamefont {G.~E.}\ \bibnamefont {Morfill}}, \bibinfo {author} {\bibfnamefont {J.}~\bibnamefont {Schablinski}}, \ and\ \bibinfo {author} {\bibfnamefont {D.}~\bibnamefont {Block}},\ }\href {\doibase 10.1209/0295-5075/100/35001} {\bibfield  {journal} {\bibinfo  {journal} {EPL (Europhysics Letters)}\ }\textbf {\bibinfo {volume} {100}},\ \bibinfo {pages} {35001} (\bibinfo {year} {2012})}\BibitemShut {NoStop}%
\bibitem [{\citenamefont {Chen}\ \emph {et~al.}(2016)\citenamefont {Chen}, \citenamefont {Dropmann}, \citenamefont {Zhang}, \citenamefont {Matthews},\ and\ \citenamefont {Hyde}}]{chen2016}%
  \BibitemOpen
  \bibfield  {author} {\bibinfo {author} {\bibfnamefont {M.}~\bibnamefont {Chen}}, \bibinfo {author} {\bibfnamefont {M.}~\bibnamefont {Dropmann}}, \bibinfo {author} {\bibfnamefont {B.}~\bibnamefont {Zhang}}, \bibinfo {author} {\bibfnamefont {L.~S.}\ \bibnamefont {Matthews}}, \ and\ \bibinfo {author} {\bibfnamefont {T.~W.}\ \bibnamefont {Hyde}},\ }\href {\doibase 10.1103/PhysRevE.94.033201} {\bibfield  {journal} {\bibinfo  {journal} {Physical Review E}\ }\textbf {\bibinfo {volume} {94}},\ \bibinfo {pages} {033201} (\bibinfo {year} {2016})}\BibitemShut {NoStop}%
\bibitem [{\citenamefont {Sütterlin}\ \emph {et~al.}(2009)\citenamefont {Sütterlin}, \citenamefont {Wysocki}, \citenamefont {Ivlev}, \citenamefont {Räth}, \citenamefont {Thomas}, \citenamefont {Rubin-Zuzic}, \citenamefont {Goedheer}, \citenamefont {Fortov}, \citenamefont {Lipaev}, \citenamefont {Molotkov}, \citenamefont {Petrov}, \citenamefont {Morfill},\ and\ \citenamefont {Löwen}}]{sutterlin2009}%
  \BibitemOpen
  \bibfield  {author} {\bibinfo {author} {\bibfnamefont {K.~R.}\ \bibnamefont {Sütterlin}}, \bibinfo {author} {\bibfnamefont {A.}~\bibnamefont {Wysocki}}, \bibinfo {author} {\bibfnamefont {A.~V.}\ \bibnamefont {Ivlev}}, \bibinfo {author} {\bibfnamefont {C.}~\bibnamefont {Räth}}, \bibinfo {author} {\bibfnamefont {H.~M.}\ \bibnamefont {Thomas}}, \bibinfo {author} {\bibfnamefont {M.}~\bibnamefont {Rubin-Zuzic}}, \bibinfo {author} {\bibfnamefont {W.~J.}\ \bibnamefont {Goedheer}}, \bibinfo {author} {\bibfnamefont {V.~E.}\ \bibnamefont {Fortov}}, \bibinfo {author} {\bibfnamefont {A.~M.}\ \bibnamefont {Lipaev}}, \bibinfo {author} {\bibfnamefont {V.~I.}\ \bibnamefont {Molotkov}}, \bibinfo {author} {\bibfnamefont {O.~F.}\ \bibnamefont {Petrov}}, \bibinfo {author} {\bibfnamefont {G.~E.}\ \bibnamefont {Morfill}}, \ and\ \bibinfo {author} {\bibfnamefont {H.}~\bibnamefont {Löwen}},\ }\href {\doibase 10.1103/PhysRevLett.102.085003} {\bibfield  {journal} {\bibinfo  {journal} {Physical Review Letters}\ }\textbf {\bibinfo
  {volume} {102}},\ \bibinfo {pages} {085003} (\bibinfo {year} {2009})}\BibitemShut {NoStop}%
\bibitem [{\citenamefont {Du}\ \emph {et~al.}(2012)\citenamefont {Du}, \citenamefont {Sütterlin}, \citenamefont {Jiang}, \citenamefont {Räth}, \citenamefont {Ivlev}, \citenamefont {Khrapak}, \citenamefont {Schwabe}, \citenamefont {Thomas}, \citenamefont {Fortov}, \citenamefont {Lipaev}, \citenamefont {Molotkov}, \citenamefont {Petrov}, \citenamefont {Malentschenko}, \citenamefont {Yurtschichin}, \citenamefont {Lonchakov},\ and\ \citenamefont {Morfill}}]{du2012}%
  \BibitemOpen
  \bibfield  {author} {\bibinfo {author} {\bibfnamefont {C.-R.}\ \bibnamefont {Du}}, \bibinfo {author} {\bibfnamefont {K.~R.}\ \bibnamefont {Sütterlin}}, \bibinfo {author} {\bibfnamefont {K.}~\bibnamefont {Jiang}}, \bibinfo {author} {\bibfnamefont {C.}~\bibnamefont {Räth}}, \bibinfo {author} {\bibfnamefont {A.~V.}\ \bibnamefont {Ivlev}}, \bibinfo {author} {\bibfnamefont {S.}~\bibnamefont {Khrapak}}, \bibinfo {author} {\bibfnamefont {M.}~\bibnamefont {Schwabe}}, \bibinfo {author} {\bibfnamefont {H.~M.}\ \bibnamefont {Thomas}}, \bibinfo {author} {\bibfnamefont {V.~E.}\ \bibnamefont {Fortov}}, \bibinfo {author} {\bibfnamefont {A.~M.}\ \bibnamefont {Lipaev}}, \bibinfo {author} {\bibfnamefont {V.~I.}\ \bibnamefont {Molotkov}}, \bibinfo {author} {\bibfnamefont {O.~F.}\ \bibnamefont {Petrov}}, \bibinfo {author} {\bibfnamefont {Y.}~\bibnamefont {Malentschenko}}, \bibinfo {author} {\bibfnamefont {F.}~\bibnamefont {Yurtschichin}}, \bibinfo {author} {\bibfnamefont {Y.}~\bibnamefont {Lonchakov}}, \ and\ \bibinfo
  {author} {\bibfnamefont {G.~E.}\ \bibnamefont {Morfill}},\ }\href {\doibase 10.1088/1367-2630/14/7/073058} {\bibfield  {journal} {\bibinfo  {journal} {New Journal of Physics}\ }\textbf {\bibinfo {volume} {14}},\ \bibinfo {pages} {073058} (\bibinfo {year} {2012})}\BibitemShut {NoStop}%
\bibitem [{\citenamefont {Chen}\ \emph {et~al.}(1992)\citenamefont {Chen}, \citenamefont {Zitter},\ and\ \citenamefont {Tao}}]{chen1992}%
  \BibitemOpen
  \bibfield  {author} {\bibinfo {author} {\bibfnamefont {T.-j.}\ \bibnamefont {Chen}}, \bibinfo {author} {\bibfnamefont {R.~N.}\ \bibnamefont {Zitter}}, \ and\ \bibinfo {author} {\bibfnamefont {R.}~\bibnamefont {Tao}},\ }\href {\doibase 10.1103/PhysRevLett.68.2555} {\bibfield  {journal} {\bibinfo  {journal} {Phys. Rev. Lett.}\ }\textbf {\bibinfo {volume} {68}},\ \bibinfo {pages} {2555} (\bibinfo {year} {1992})}\BibitemShut {NoStop}%
\bibitem [{\citenamefont {Ivlev}\ \emph {et~al.}(2008)\citenamefont {Ivlev}, \citenamefont {Morfill}, \citenamefont {Thomas}, \citenamefont {Räth}, \citenamefont {Joyce}, \citenamefont {Huber}, \citenamefont {Kompaneets}, \citenamefont {Fortov}, \citenamefont {Lipaev}, \citenamefont {Molotkov}, \citenamefont {Reiter}, \citenamefont {Turin},\ and\ \citenamefont {Vinogradov}}]{ivlev2008}%
  \BibitemOpen
  \bibfield  {author} {\bibinfo {author} {\bibfnamefont {A.~V.}\ \bibnamefont {Ivlev}}, \bibinfo {author} {\bibfnamefont {G.~E.}\ \bibnamefont {Morfill}}, \bibinfo {author} {\bibfnamefont {H.~M.}\ \bibnamefont {Thomas}}, \bibinfo {author} {\bibfnamefont {C.}~\bibnamefont {Räth}}, \bibinfo {author} {\bibfnamefont {G.}~\bibnamefont {Joyce}}, \bibinfo {author} {\bibfnamefont {P.}~\bibnamefont {Huber}}, \bibinfo {author} {\bibfnamefont {R.}~\bibnamefont {Kompaneets}}, \bibinfo {author} {\bibfnamefont {V.~E.}\ \bibnamefont {Fortov}}, \bibinfo {author} {\bibfnamefont {A.~M.}\ \bibnamefont {Lipaev}}, \bibinfo {author} {\bibfnamefont {V.~I.}\ \bibnamefont {Molotkov}}, \bibinfo {author} {\bibfnamefont {T.}~\bibnamefont {Reiter}}, \bibinfo {author} {\bibfnamefont {M.}~\bibnamefont {Turin}}, \ and\ \bibinfo {author} {\bibfnamefont {P.}~\bibnamefont {Vinogradov}},\ }\href {\doibase 10.1103/PhysRevLett.100.095003} {\bibfield  {journal} {\bibinfo  {journal} {Physical Review Letters}\ }\textbf {\bibinfo {volume} {100}},\
  \bibinfo {pages} {095003} (\bibinfo {year} {2008})}\BibitemShut {NoStop}%
\bibitem [{\citenamefont {Joshi}\ \emph {et~al.}(2023)\citenamefont {Joshi}, \citenamefont {Pustylnik}, \citenamefont {Thoma}, \citenamefont {Thomas},\ and\ \citenamefont {Schwabe}}]{joshi2023}%
  \BibitemOpen
  \bibfield  {author} {\bibinfo {author} {\bibfnamefont {E.}~\bibnamefont {Joshi}}, \bibinfo {author} {\bibfnamefont {M.~Y.}\ \bibnamefont {Pustylnik}}, \bibinfo {author} {\bibfnamefont {M.~H.}\ \bibnamefont {Thoma}}, \bibinfo {author} {\bibfnamefont {H.~M.}\ \bibnamefont {Thomas}}, \ and\ \bibinfo {author} {\bibfnamefont {M.}~\bibnamefont {Schwabe}},\ }\href {\doibase 10.1103/PhysRevResearch.5.L012030} {\bibfield  {journal} {\bibinfo  {journal} {Phys. Rev. Res.}\ }\textbf {\bibinfo {volume} {5}},\ \bibinfo {pages} {L012030} (\bibinfo {year} {2023})}\BibitemShut {NoStop}%
\bibitem [{\citenamefont {Matthews}\ \emph {et~al.}(2021)\citenamefont {Matthews}, \citenamefont {Vermillion}, \citenamefont {Hartmann}, \citenamefont {Rosenberg}, \citenamefont {Rostami}, \citenamefont {Kostadinova}, \citenamefont {Hyde}, \citenamefont {Pustylnik}, \citenamefont {Lipaev}, \citenamefont {Usachev}, \citenamefont {Zobnin}, \citenamefont {Thoma}, \citenamefont {Petrov}, \citenamefont {Thomas},\ and\ \citenamefont {Novitskiy}}]{matthews2021}%
  \BibitemOpen
  \bibfield  {author} {\bibinfo {author} {\bibfnamefont {L.}~\bibnamefont {Matthews}}, \bibinfo {author} {\bibfnamefont {K.}~\bibnamefont {Vermillion}}, \bibinfo {author} {\bibfnamefont {P.}~\bibnamefont {Hartmann}}, \bibinfo {author} {\bibfnamefont {M.}~\bibnamefont {Rosenberg}}, \bibinfo {author} {\bibfnamefont {S.}~\bibnamefont {Rostami}}, \bibinfo {author} {\bibfnamefont {E.}~\bibnamefont {Kostadinova}}, \bibinfo {author} {\bibfnamefont {T.}~\bibnamefont {Hyde}}, \bibinfo {author} {\bibfnamefont {M.}~\bibnamefont {Pustylnik}}, \bibinfo {author} {\bibfnamefont {A.}~\bibnamefont {Lipaev}}, \bibinfo {author} {\bibfnamefont {A.}~\bibnamefont {Usachev}}, \bibinfo {author} {\bibfnamefont {A.}~\bibnamefont {Zobnin}}, \bibinfo {author} {\bibfnamefont {M.}~\bibnamefont {Thoma}}, \bibinfo {author} {\bibfnamefont {O.}~\bibnamefont {Petrov}}, \bibinfo {author} {\bibfnamefont {H.}~\bibnamefont {Thomas}}, \ and\ \bibinfo {author} {\bibfnamefont {O.}~\bibnamefont {Novitskiy}},\ }\href {\doibase
  10.1017/S0022377821001215} {\bibfield  {journal} {\bibinfo  {journal} {Journal of Plasma Physics}\ }\textbf {\bibinfo {volume} {87}},\ \bibinfo {pages} {905870618} (\bibinfo {year} {2021})}\BibitemShut {NoStop}%
\bibitem [{\citenamefont {Vermillion}\ \emph {et~al.}(2022)\citenamefont {Vermillion}, \citenamefont {Sanford}, \citenamefont {Matthews}, \citenamefont {Hartmann}, \citenamefont {Rosenberg}, \citenamefont {Kostadinova}, \citenamefont {Carmona-Reyes}, \citenamefont {Hyde}, \citenamefont {Lipaev}, \citenamefont {Usachev}, \citenamefont {Zobnin}, \citenamefont {Petrov}, \citenamefont {Thoma}, \citenamefont {Pustylnik}, \citenamefont {Thomas},\ and\ \citenamefont {Ovchinin}}]{vermillion2022}%
  \BibitemOpen
  \bibfield  {author} {\bibinfo {author} {\bibfnamefont {K.}~\bibnamefont {Vermillion}}, \bibinfo {author} {\bibfnamefont {D.}~\bibnamefont {Sanford}}, \bibinfo {author} {\bibfnamefont {L.}~\bibnamefont {Matthews}}, \bibinfo {author} {\bibfnamefont {P.}~\bibnamefont {Hartmann}}, \bibinfo {author} {\bibfnamefont {M.}~\bibnamefont {Rosenberg}}, \bibinfo {author} {\bibfnamefont {E.}~\bibnamefont {Kostadinova}}, \bibinfo {author} {\bibfnamefont {J.}~\bibnamefont {Carmona-Reyes}}, \bibinfo {author} {\bibfnamefont {T.}~\bibnamefont {Hyde}}, \bibinfo {author} {\bibfnamefont {A.~M.}\ \bibnamefont {Lipaev}}, \bibinfo {author} {\bibfnamefont {A.~D.}\ \bibnamefont {Usachev}}, \bibinfo {author} {\bibfnamefont {A.~V.}\ \bibnamefont {Zobnin}}, \bibinfo {author} {\bibfnamefont {O.~F.}\ \bibnamefont {Petrov}}, \bibinfo {author} {\bibfnamefont {M.~H.}\ \bibnamefont {Thoma}}, \bibinfo {author} {\bibfnamefont {M.~Y.}\ \bibnamefont {Pustylnik}}, \bibinfo {author} {\bibfnamefont {H.~M.}\ \bibnamefont {Thomas}}, \ and\ \bibinfo
  {author} {\bibfnamefont {A.}~\bibnamefont {Ovchinin}},\ }\href {\doibase 10.1063/5.0075261} {\bibfield  {journal} {\bibinfo  {journal} {Physics of Plasmas}\ }\textbf {\bibinfo {volume} {29}},\ \bibinfo {pages} {023701} (\bibinfo {year} {2022})}\BibitemShut {NoStop}%
\bibitem [{\citenamefont {Ivlev}\ \emph {et~al.}(2011)\citenamefont {Ivlev}, \citenamefont {Thoma}, \citenamefont {R\"ath}, \citenamefont {Joyce},\ and\ \citenamefont {Morfill}}]{ivlev2011}%
  \BibitemOpen
  \bibfield  {author} {\bibinfo {author} {\bibfnamefont {A.~V.}\ \bibnamefont {Ivlev}}, \bibinfo {author} {\bibfnamefont {M.~H.}\ \bibnamefont {Thoma}}, \bibinfo {author} {\bibfnamefont {C.}~\bibnamefont {R\"ath}}, \bibinfo {author} {\bibfnamefont {G.}~\bibnamefont {Joyce}}, \ and\ \bibinfo {author} {\bibfnamefont {G.~E.}\ \bibnamefont {Morfill}},\ }\href {\doibase 10.1103/PhysRevLett.106.155001} {\bibfield  {journal} {\bibinfo  {journal} {Phys. Rev. Lett.}\ }\textbf {\bibinfo {volume} {106}},\ \bibinfo {pages} {155001} (\bibinfo {year} {2011})}\BibitemShut {NoStop}%
\bibitem [{\citenamefont {Schmidt}\ and\ \citenamefont {Hyde}(2020)}]{schmidt2020}%
  \BibitemOpen
  \bibfield  {author} {\bibinfo {author} {\bibfnamefont {J.}~\bibnamefont {Schmidt}}\ and\ \bibinfo {author} {\bibfnamefont {T.}~\bibnamefont {Hyde}},\ }\href@noop {} {\bibfield  {journal} {\bibinfo  {journal} {Review of Scientific Instruments}\ }\textbf {\bibinfo {volume} {91}} (\bibinfo {year} {2020})}\BibitemShut {NoStop}%
\bibitem [{\citenamefont {Matthews}\ \emph {et~al.}(2020)\citenamefont {Matthews}, \citenamefont {Sanford}, \citenamefont {Kostadinova}, \citenamefont {Ashrafi}, \citenamefont {Guay},\ and\ \citenamefont {Hyde}}]{matthews2020}%
  \BibitemOpen
  \bibfield  {author} {\bibinfo {author} {\bibfnamefont {L.~S.}\ \bibnamefont {Matthews}}, \bibinfo {author} {\bibfnamefont {D.~L.}\ \bibnamefont {Sanford}}, \bibinfo {author} {\bibfnamefont {E.~G.}\ \bibnamefont {Kostadinova}}, \bibinfo {author} {\bibfnamefont {K.~S.}\ \bibnamefont {Ashrafi}}, \bibinfo {author} {\bibfnamefont {E.}~\bibnamefont {Guay}}, \ and\ \bibinfo {author} {\bibfnamefont {T.~W.}\ \bibnamefont {Hyde}},\ }\href {\doibase 10.1063/1.5124246} {\bibfield  {journal} {\bibinfo  {journal} {Physics of Plasmas}\ }\textbf {\bibinfo {volume} {27}},\ \bibinfo {pages} {023703} (\bibinfo {year} {2020})}\BibitemShut {NoStop}%
\bibitem [{\citenamefont {Ivlev}\ \emph {et~al.}(2010)\citenamefont {Ivlev}, \citenamefont {Brandt}, \citenamefont {Morfill}, \citenamefont {Rath}, \citenamefont {Thomas}, \citenamefont {Joyce}, \citenamefont {Fortov}, \citenamefont {Lipaev}, \citenamefont {Molotkov},\ and\ \citenamefont {Petrov}}]{ivlev2010}%
  \BibitemOpen
  \bibfield  {author} {\bibinfo {author} {\bibfnamefont {A.~V.}\ \bibnamefont {Ivlev}}, \bibinfo {author} {\bibfnamefont {P.~C.}\ \bibnamefont {Brandt}}, \bibinfo {author} {\bibfnamefont {G.~E.}\ \bibnamefont {Morfill}}, \bibinfo {author} {\bibfnamefont {C.}~\bibnamefont {Rath}}, \bibinfo {author} {\bibfnamefont {H.~M.}\ \bibnamefont {Thomas}}, \bibinfo {author} {\bibfnamefont {G.}~\bibnamefont {Joyce}}, \bibinfo {author} {\bibfnamefont {V.~E.}\ \bibnamefont {Fortov}}, \bibinfo {author} {\bibfnamefont {A.~M.}\ \bibnamefont {Lipaev}}, \bibinfo {author} {\bibfnamefont {V.~I.}\ \bibnamefont {Molotkov}}, \ and\ \bibinfo {author} {\bibfnamefont {O.~F.}\ \bibnamefont {Petrov}},\ }\href {\doibase 10.1109/TPS.2009.2037716} {\bibfield  {journal} {\bibinfo  {journal} {IEEE Transactions on Plasma Science}\ }\textbf {\bibinfo {volume} {38}},\ \bibinfo {pages} {733} (\bibinfo {year} {2010})}\BibitemShut {NoStop}%
\bibitem [{\citenamefont {Pustylnik}\ \emph {et~al.}(2016)\citenamefont {Pustylnik}, \citenamefont {Fink}, \citenamefont {Nosenko}, \citenamefont {Antonova}, \citenamefont {Hagl}, \citenamefont {Thomas}, \citenamefont {Zobnin}, \citenamefont {Lipaev}, \citenamefont {Usachev}, \citenamefont {Molotkov}, \citenamefont {Petrov}, \citenamefont {Fortov}, \citenamefont {Rau}, \citenamefont {Deysenroth}, \citenamefont {Albrecht}, \citenamefont {Kretschmer}, \citenamefont {Thoma}, \citenamefont {Morfill}, \citenamefont {Seurig}, \citenamefont {Stettner}, \citenamefont {Alyamovskaya}, \citenamefont {Orr}, \citenamefont {Kufner}, \citenamefont {Lavrenko}, \citenamefont {Padalka}, \citenamefont {Serova}, \citenamefont {Samokutyayev},\ and\ \citenamefont {Christoforetti}}]{pustylnik2016}%
  \BibitemOpen
  \bibfield  {author} {\bibinfo {author} {\bibfnamefont {M.~Y.}\ \bibnamefont {Pustylnik}}, \bibinfo {author} {\bibfnamefont {M.~A.}\ \bibnamefont {Fink}}, \bibinfo {author} {\bibfnamefont {V.}~\bibnamefont {Nosenko}}, \bibinfo {author} {\bibfnamefont {T.}~\bibnamefont {Antonova}}, \bibinfo {author} {\bibfnamefont {T.}~\bibnamefont {Hagl}}, \bibinfo {author} {\bibfnamefont {H.~M.}\ \bibnamefont {Thomas}}, \bibinfo {author} {\bibfnamefont {A.~V.}\ \bibnamefont {Zobnin}}, \bibinfo {author} {\bibfnamefont {A.~M.}\ \bibnamefont {Lipaev}}, \bibinfo {author} {\bibfnamefont {A.~D.}\ \bibnamefont {Usachev}}, \bibinfo {author} {\bibfnamefont {V.~I.}\ \bibnamefont {Molotkov}}, \bibinfo {author} {\bibfnamefont {O.~F.}\ \bibnamefont {Petrov}}, \bibinfo {author} {\bibfnamefont {V.~E.}\ \bibnamefont {Fortov}}, \bibinfo {author} {\bibfnamefont {C.}~\bibnamefont {Rau}}, \bibinfo {author} {\bibfnamefont {C.}~\bibnamefont {Deysenroth}}, \bibinfo {author} {\bibfnamefont {S.}~\bibnamefont {Albrecht}}, \bibinfo {author}
  {\bibfnamefont {M.}~\bibnamefont {Kretschmer}}, \bibinfo {author} {\bibfnamefont {M.~H.}\ \bibnamefont {Thoma}}, \bibinfo {author} {\bibfnamefont {G.~E.}\ \bibnamefont {Morfill}}, \bibinfo {author} {\bibfnamefont {R.}~\bibnamefont {Seurig}}, \bibinfo {author} {\bibfnamefont {A.}~\bibnamefont {Stettner}}, \bibinfo {author} {\bibfnamefont {V.~A.}\ \bibnamefont {Alyamovskaya}}, \bibinfo {author} {\bibfnamefont {A.}~\bibnamefont {Orr}}, \bibinfo {author} {\bibfnamefont {E.}~\bibnamefont {Kufner}}, \bibinfo {author} {\bibfnamefont {E.~G.}\ \bibnamefont {Lavrenko}}, \bibinfo {author} {\bibfnamefont {G.~I.}\ \bibnamefont {Padalka}}, \bibinfo {author} {\bibfnamefont {E.~O.}\ \bibnamefont {Serova}}, \bibinfo {author} {\bibfnamefont {A.~M.}\ \bibnamefont {Samokutyayev}}, \ and\ \bibinfo {author} {\bibfnamefont {S.}~\bibnamefont {Christoforetti}},\ }\href {\doibase 10.1063/1.4962696} {\bibfield  {journal} {\bibinfo  {journal} {Review of Scientific Instruments}\ }\textbf {\bibinfo {volume} {87}},\ \bibinfo {pages}
  {093505} (\bibinfo {year} {2016})}\BibitemShut {NoStop}%
\bibitem [{\citenamefont {Vermillion}\ \emph {et~al.}(2024)\citenamefont {Vermillion}, \citenamefont {Banka}, \citenamefont {Mendoza}, \citenamefont {Wyatt}, \citenamefont {Matthews},\ and\ \citenamefont {Hyde}}]{vermillion2024}%
  \BibitemOpen
  \bibfield  {author} {\bibinfo {author} {\bibfnamefont {K.}~\bibnamefont {Vermillion}}, \bibinfo {author} {\bibfnamefont {R.}~\bibnamefont {Banka}}, \bibinfo {author} {\bibfnamefont {A.}~\bibnamefont {Mendoza}}, \bibinfo {author} {\bibfnamefont {B.}~\bibnamefont {Wyatt}}, \bibinfo {author} {\bibfnamefont {L.}~\bibnamefont {Matthews}}, \ and\ \bibinfo {author} {\bibfnamefont {T.}~\bibnamefont {Hyde}},\ }\href@noop {} {\bibfield  {journal} {\bibinfo  {journal} {Physics of Plasmas}\ }\textbf {\bibinfo {volume} {31}} (\bibinfo {year} {2024})}\BibitemShut {NoStop}%
\end{thebibliography}%

\end{document}